\PassOptionsToPackage{dvipsnames}{xcolor} 
\documentclass[acmsmall,screen]{acmart}
\usepackage{booktabs}   
\usepackage{subcaption} 

\usepackage[inference]{semantic}
\usepackage{amsmath}\allowdisplaybreaks
\usepackage{mathpartir}
\usepackage{natbib}
\usepackage{amsthm}
\usepackage{stmaryrd} 
\usepackage{xspace}
\usepackage{latexsym}
\usepackage{hyperref}
\usepackage{ifthen}
\usepackage{mathtools}
\usepackage{color}
\usepackage{listings}
\usepackage{verbatim}
\usepackage[colorinlistoftodos]{todonotes}
\usepackage{tikz}
\usetikzlibrary{positioning,shadows,arrows,calc,backgrounds,fit,shapes,shapes.multipart,decorations.pathreplacing,shapes.misc,patterns}
\usetikzlibrary{3d}
\tikzset{myimpl/.style={,double,-implies,double equal sign distance}}
\usepackage{tikzscale}
\usepackage{xspace}
\usepackage[scaled=.83]{beramono}
\usepackage{epigraph}
\usepackage[capitalize]{cleveref}
\usepackage{booktabs}
\usepackage{float}
\usepackage{etoolbox}
\usepackage{wrapfig}
\usepackage{scalerel}

\usepackage{bm}
\usepackage{nameref}
\usepackage{bussproofs}
\usepackage{textcomp}
\usepackage{textalpha}
\usepackage{paralist}
\usepackage{xargs}
\usepackage{xparse}

\hypersetup{ pdfpagemode=UseOutlines, colorlinks=true, linkcolor=OliveGreen, citecolor=blue }


\DeclareMathAlphabet{\mathsf}{OT1}{cmss}{m}{n}
\SetMathAlphabet{\mathsf}{bold}{OT1}{cmss}{bx}{n}

\newcommand{\papertitle}[0]{\mswasm: Soundly Enforcing Memory-Safe Execution of Unsafe Code}

\newcommandx{\MV}[2][1=]{\todo[color=yellow!30,#1]{MV: #2}}
\newcommandx{\mv}[2][1=]{\todo[color=yellow!30,#1]{MV: #2}}
\newcommandx{\cd}[2][1=]{\todo[color=green!30,#1]{CD: #2}}
\newcommandx{\jb}[2][1=]{\todo[color=purple!30,#1]{JB: #2}}
\newcommandx{\bp}[2][1=]{\todo[color=orange!30,#1]{BP: #2}}
\newcommandx{\am}[2][1=]{\todo[color=cyan!30,#1]{AM: #2}}

\newcommand{\mi}[1]{\ensuremath{\mathit{#1}}}

\newcommand{\mtt}[1]{\ensuremath{\mathtt{#1}}}
\newcommand{\mf}[1]{\ensuremath{\mathbf{#1}}}

\newcommand{\mc}[1]{\ensuremath{\mathcal{#1}}}
\newcommand{\ms}[1]{\ensuremath{\mathsf{#1}}}
\newcommand{\mb}[1]{\ensuremath{\mathbb{#1}}}

\newcommand{\isdef}[0]{\ensuremath{\mathrel{\overset{\makebox[0pt]{\mbox{\normalfont\tiny\sffamily def}}}{=}}}}

\newcommand\bnfdef{\ensuremath{\mathrel{::=}}}

\newcommand{\OB}[1]{\ensuremath{\overline{#1}}}

\newcommand*{\QEDA}{\hfill\ensuremath{\blacksquare}}%

\AtEndEnvironment{problem}{\null\hfill\QEDA}
\AtEndEnvironment{example}{}

\Crefname{lstlisting}{Listing}{Listings}
\Crefname{problem}{Problem}{Problems}

\Crefname{equation}{Rule}{Rules}

\newenvironment{proofsketch}{\trivlist\item[]\emph{Proof (Sketch)}.\xspace}{\unskip\nobreak\hskip 1em plus 1fil\nobreak$\Box$\parfillskip=0pt\endtrivlist}
\newcommand{\compskel}[3]{\ensuremath{\bl{\left\llbracket \src{#1} \right\rrbracket^{#2}_{#3}}}}
\newcommand{\compskelnocolor}[3]{\ensuremath{\bl{\left\llbracket {#1} \right\rrbracket^{#2}_{#3}}}}
\newcommand{\comp}[1]{\compskel{\bl{#1}}{}{}}

\newcommand{\funname}[1]{\mtt{#1}}
\newcommand{\fun}[2]{\ensuremath{\bl{\funname{#1}\left(#2\right)}}\xspace}

\newcommand{\compc}[3]{\compskelnocolor{#2 }{\src{exp}}{} = #3}
\newcommand{\compct}[2]{\comp{#1} = #2}

\newcommand{\composition}{\ensuremath{\circ}}
\newcommand{\compose}[2]{\ensuremath{#1 \composition #2}}

\newcommand{\hequiv}[3]{\ensuremath{#1 \sim_{#2} #3}}

\newcommand{\compd}[3]{\ensuremath{ #2 \sim_{\crossdelta{}} #3}}

\newcommand{\srcms}[1]{\src{\texttt{MS}(#1)}}
\newcommand{\trgms}[1]{\trg{\texttt{MS}(#1)}}

\newcommand{\relatesrctrgtr}[3]{\ensuremath{ \src{#1}~=_{#2} ~\trg{#3}   }}

\newcommand{\traceeq}[4]{\ensuremath{#1 =_{#3} #4}}
\newcommand{\traceneq}[4]{\ensuremath{#1 \neq_{#3} #4}}
\newcommand{\srcdelta}[1]{\src{\delta{#1}}}
\newcommand{\trgdelta}[1]{\trg{\trgb{\delta}{#1}} }
\newcommand{\crossdelta}[1]{\com{\delta{#1}} }

\newcommand{\contextletter}[0]{C}
\newcommand{\ctx}[0]{\ensuremath{\contextletter}}

\newcommand{\add}{\ensuremath{\trg{handle.add}}}

\newcommand{\neutcol}[0]{black}
\newcommand{\stlccol}[0]{RoyalBlue}
\newcommand{\ulccol}[0]{RedOrange}

\newcommand{\commoncol}[0]{black}    

\newcommand{\col}[2]{\ensuremath{{\color{#1}{#2}}}}

\newcommand{\src}[1]{\ms{\col{\stlccol}{#1}}}
\newcommand{\trg}[1]{\mf{\col{\ulccol }{#1}}}
\newcommand{\trgb}[1]{\ensuremath{\bm{\col{\ulccol }{#1}}}}

\newcommand{\bl}[1]{\col{\neutcol }{#1}}
\newcommand{\com}[1]{\mi{\col{\commoncol }{#1}}}

\newcounter{typerule}
\crefname{typerule}{rule}{rules}

\newcommand{\srule}[1]{\textsf{(#1)}}

\newcommand{\typeruleInt}[5]{
	\def\thetyperule{#1}%
	\refstepcounter{typerule}%
	\label{tr:#4}%
  \ensuremath{\begin{array}{c}#5 \inference{#2}{#3}\end{array}}
}
\newcommand{\typerule}[4]{
  \typeruleInt{\srule{#1}}{#2}{#3}{#4}{\scriptsize\srule{#1} \\      }
}

\newcommand{\trule}[3]{\ensuremath{#1 \vdash #2 : #3}}

\newcommand{\strule}[3]{\trule{\src{#1}}{\src{#2}}{\src{#3}}}

\newcommand{\struleprog}[1]{\ensuremath{\vdash \src{#1} : \src{wt}}}

\pgfdeclarelayer{background}
\pgfdeclarelayer{veryback}
\pgfdeclarelayer{veryback2}
\pgfdeclarelayer{veryback3}
\pgfdeclarelayer{back2}
\pgfdeclarelayer{foreground}
\pgfsetlayers{veryback3,veryback2,veryback,background,back2,main,foreground}

\newcommand{\myfig}[3]{\begin{figure} [!t]
#1
\caption{\label{fig:#2}#3}
\end{figure}}

\newcommand{\eg}[0]{e.g.,\xspace}

\newcommand{\myparagraph}[1]{\smallskip\noindent\textbf{#1}\xspace}

\newcommand{\srcconf}[1]{\src{\langle#1\rangle}}

\newcommand{\srcconfig}[3]{\src{\langle #1, #2, #3 \rangle}}

\newcommand{\srcstepprimconfig}[3]{\ensuremath{\src{M \vdash #1 \stepredx{#2} #3}} }

\newcommand{\srcastep}[3]{\src{#1\aarrow{#2}#3}}

\newcommand{\funt}[2]{\trg{{#1}\left({#2}\right)}}

\newcommand{\abs}[1]{\ensuremath{\lvert#1\rvert}}

\newcommand{\iftt}[2]{\trg{branch}~#1~#2~}

\newcommand{\conf}[1]{\trg{\langle#1\rangle}}

\newcommand{\stackcat}{\ensuremath{:}}
\newcommand{\stackapp}{\ensuremath{+{\mkern-16mu}+}}

\newcommand{\xrightarrowdbl}[2][]{%
  \xrightarrow[#1]{#2}\mathrel{\mkern-14mu}\rightarrow
}

\newcommand{\xtoudbl}[3]{\ensuremath{~\mathrel{\xrightarrowdbl{~#1~}^{#2}_{#3}}}}

\newcommand{\fdef}{\Phi}

\newcommand{\trgfuncs}{\trg{\stk{\fdef}}}

\newcommand{\wstodbl}[4]{ \trg{\ensuremath{\trgfuncs \vdash #1\xtoudbl{#2}{}{#3}#4}}}

\newcommand{\wstof}[3]{\wstodbl{#1}{#2}{}{#3}}

\usepackage{mathtools}
\usepackage{tikz-cd}
\usetikzlibrary{decorations.pathmorphing}

\newcommand{\aarrow}[1]{\xhookrightarrow{#1}}
\newcommand{\astep}[3]{\trg{#1\aarrow{#2}#3}}

\newcommand{\bopt}{\trg{\trgb{\tau}.\bop}}
\newcommand{\loadt}{\trg{\trgb{\tau}.load}}
\newcommand{\storet}{\trg{\trgb{\tau}.store~}}

\newcommand{\hloadtext}{segload}
\newcommand{\hstoretext}{segstore}
\newcommand{\hslicetext}{slice}
\newcommand{\newsttext}{segalloc}

\newcommand{\handle}{\trg{handle}\xspace}
\newcommand{\handles}{{\handle}s\xspace}
\newcommand{\ittw}{i32}
\newcommand{\hload}{\trg{\trgb{\tau}.\hloadtext}\xspace}
\newcommand{\hstore}{\trg{\trgb{\tau}.\hstoretext}\xspace}
\newcommand{\hslice}{\trg{\hslicetext}}

\newcommand{\hstoretau}[1]{\trg{{#1}.segment\_store}}
\newcommand{\newst}{\trg{\newsttext}} 

\newcommand{\freest}{\trg{segfree}}

\newcommand{\writet}[1]{\trg{write_{\trgb{\tau}}\left({#1}\right)}}
\newcommand{\readt}[1]{\trg{read_{\trgb{\tau}}\left({#1}\right)}}

\newcommand{\hl}[1]{\colorbox{yellow}{#1}}
\newcounter{line}

\newcommand{\stepredx}[1]{\ensuremath{~\mathrel{\xrightarrowdbl{~#1~}\phantom{.\!\!}}~}}

\definecolor{mygreen}{rgb}{0,0.6,0}
\definecolor{mygray}{rgb}{0.5,0.5,0.5}
\definecolor{mymauve}{rgb}{0.58,0,0.82}

\lstdefinelanguage{Java} 
{morekeywords={abstract, all, and, as, assert, but, disj, else, exactly, extends, fact, for, fun, iden, if, iff, implies, in, Int, int, let, lone, module, no, none, not, one, open, or, part, pred, run, seq, set, sig, some, sum, then, univ, package, class, public, private, null, return, new, interface, extern, object, implements, System, static, super, try , catch, throw, throws, Unit, var, val, of, principal, trust},
sensitive=true,
keywordstyle=\bfseries\color{green!40!black},
commentstyle=\itshape\color{purple!40!black},
morecomment=[l][\small\itshape\color{purple!40!black}]{//},
identifierstyle=\color{blue},
stringstyle=\color{orange},
basicstyle=\small,
basicstyle={\small\ttfamily},
numbers=left,
numberstyle=\tiny\color{mygray},
tabsize=2,
numbersep=3pt,
breaklines=true,
lineskip=-2pt,
stepnumber=1,
captionpos=b,
breaklines=true,
breakatwhitespace=false,
showspaces=false,
showtabs=false,
float=!h,
columns=fullflexible,escapeinside={(*@}{@*)},
moredelim=**[is][\color{red!60}]{@}{@},
literate={->}{{$\to$}}1 {^}{{$\mspace{-3mu}\widehat{\quad}\mspace{-3mu}$}}1
{<}{$<$ }2 {>}{$>$ }2 {>=}{$\geq$ }2 {=<}{$\leq$ }2
{<:}{{$<\mspace{-3mu}:$}}2 {:>}{{$:\mspace{-3mu}>$}}2
{=>}{{$\Rightarrow$ }}2 
{<=>}{{$\Leftrightarrow$ }}2 
{\~}{{$\mspace{-3mu}\widetilde{\quad}\mspace{-3mu}$}}1
{!=}{$\neq$ }2 {*}{${}^{\ast}$}1 
{\#}{$\#$}1
}

\lstdefinelanguage{Asm}
{morekeywords={abstract, all, and, as, assert, but, check, disj, else, exactly, extends, fact, for, fun, iden, if, iff, implies, in, Int, int, let, lone, module, no, none, not, one, open, or, part, pred, run, seq, set, sig, some, sum, then, univ, package, class, public, private, null, return, new, interface, extern, object, implements, System, static, super, try , catch, throw, throws, Unit, var, val, principal, trust, label, load, add, addi, into, test},
sensitive=true,
identifierstyle=\color{black},
keywordstyle=\bfseries,
commentstyle=\itshape\color{purple!40!black},
morecomment=[l][\small\itshape\color{purple!40!black}]{//},
stringstyle=\color{orange},
basicstyle=\small,
basicstyle={\small},
numbers=left,
numberstyle=\tiny\color{mygray},
tabsize=2,
numbersep=3pt,
breaklines=true,
lineskip=-2pt,
stepnumber=1,
captionpos=b,
breaklines=true,
breakatwhitespace=false,
showspaces=false,
showtabs=false,
float=!h,
columns=fullflexible,escapeinside={(*@}{@*)},
moredelim=**[is][\color{red!60}]{@}{@},
literate={->}{{$\to$}}1 {^}{{$\mspace{-3mu}\widehat{\quad}\mspace{-3mu}$}}1
{<}{$<$ }2 {>}{$>$ }2 {>=}{$\geq$ }2 {=<}{$\leq$ }2
{<:}{{$<\mspace{-3mu}:$}}2 {:>}{{$:\mspace{-3mu}>$}}2
{=>}{{$\Rightarrow$ }}2 
{<=>}{{$\Leftrightarrow$ }}2 
{\~}{{$\mspace{-3mu}\widetilde{\quad}\mspace{-3mu}$}}1
{!=}{$\neq$ }2 {*}{${}^{\ast}$}1 
{\#}{$\#$}1
}
\DeclareMathOperator\ceq{\ensuremath{\mathrel{\simeq_{\mi{ctx}}}}}

\def\teqaux#1{\vcenter{\hbox{\ooalign{\hfil
       \raise6pt \hbox{\scriptsize{T}}\hfil\cr\hfil
       $=$}}}}

\def\ceqwaux#1{\vcenter{\hbox{\ooalign{\hfil
       \raise6pt \hbox{\scriptsize{w-b}}\hfil\cr\hfil
       $\ceq$}}}}

\def\praux#1{\vcenter{\hbox{\ooalign{\hfil
       \raise4pt \hbox{$\subset$}\hfil\cr\hfil
       $\sim$}}}}

\newcommand{\alphaseq}[0]{\OB{\alpha}\xspace}
\newcommand{\alphaseqp}[0]{\OB{\alpha'}\xspace}
\newcommand{\alss}[0]{\src{\alphaseq}\xspace}
\newcommand{\alssp}[0]{\src{\alphaseqp}\xspace}

\newcommand{\alst}[0]{\trgb{\alphaseq}\xspace}

\newcommand{\alsc}[0]{\com{\alphaseq}\xspace}

\newcommand{\as}[0]{\src{\alpha}\xspace}
\newcommand{\at}[0]{\trgb{\alpha}\xspace}
\newcommand{\ac}[0]{\com{\alpha}\xspace}

\newcommand{\rd}[2]{\traceform{\funname{read}(#1^{#2})}}
\newcommand{\wrt}[2]{\traceform{\funname{write}(#1^{#2})}}

\newcommand{\psrcrd}[2]{\src{read_{#1}(#2)}}
\newcommand{\psrcwrt}[2]{\src{write_#1(#2)}}

\newcommand{\sallocgen}[4]{\traceform{\funname{alloc}(#1, #2^{#3}, #4)}} 

\newcommand{\frees}[1]{\traceform{{free}(#1)}}

\newcommand{\sfree}[2]{\traceform{\fun{sfree}{#1^{#2}}}}
\newcommand{\fresh}[1]{\fun{fresh}{#1}}

\newcommand{\rdt}[1]{\traceform{\funt{read_{\trgb{\tau}}\!}{#1}}}
\newcommand{\wrtt}[1]{\traceform{\funt{write_{\trgb{\tau}}\!}{#1}}}
\newcommand{\psalloct}[2]{\traceform{\funt{salloc}{#1, #2}}} 
\newcommand{\trapt}{\traceform{\trg{trap}}}

\NewDocumentCommand\mto{omom}{%
    \ensuremath{\IfValueT{#1}{#1\ \vdash}#2\xrightarrow{\IfValueT{#3}{#3}} #4}}

\NewDocumentCommand\notmto{omom}{%
    \ensuremath{\IfValueT{#1}{#1\ \vdash}#2 \not \xrightarrow{\IfValueT{#3}{#3}} #4}}

\newcommand{\sz}[1]{\fun{sz}{#1}}

\theoremstyle{definition}

\newtheorem{definition}{Definition}
\newtheorem{theorem}{Theorem}

\newtheorem{corollary}{Corollary}
\Crefname{corollary}{Corollary}{Corollaries}
\Crefname{informal}{Definition}{Definition}
\Crefname{assumption}{Assumption}{Assumptions}
\crefname{assumption}{Assumption}{Assumptions}
\Crefname{property}{Property}{Properties}
\crefname{property}{Property}{Properties}

\Crefname{paragraph}{Section}{Sections}

\newcommand{\op}[0]{\ensuremath{\oplus}}
\newcommand{\bop}[0]{\ensuremath{\otimes}}

\newcommand{\struct}[1]{struct~#1}

\newcommand{\aptr}[4]{#1^{(#2, #3, #4, \tid)}}

\newcommand{\ptrc}[1]{ptr\, #1}
\newcommand{\arr}[2]{array~#2}

\newcommand{\pmalloct}[2]{malloc(#1,#2)}
\newcommand{\pmalloc}[1]{malloc(#1)}
\newcommand{\freesrc}[1]{free(#1)}
\newcommand{\lkup}[2]{\&#1\to#2}
\newcommand{\drf}[1]{*#1}
\newcommand{\asgn}[2]{*#1:=#2}
\newcommand{\varasgn}[2]{#1:=#2}

\newcommand{\seq}[2]{\ensuremath{#1 ; #2} }

\newcommand{\vdashm}[0]{\vdash}

\newcommand{\letcall}[4]{#1: ={call}~#2\left(#3\right)}

\newcommand{\ifte}[3]{{if}~#1~{then}~#2~{else}~#3}

\newcommand{\srcvar}{\src{var}}

\newcommand{\wasm}[0]{\text{Wasm}\xspace}
\newcommand{\mswasm}[0]{\text{MSWasm}\xspace}

\newcommand{\rwasm}[0]{\text{rWasm}\xspace}
\newcommand{\rwasmdefault}[0]{\rwasm{}$_{STH}$\xspace}
\newcommand{\rwasmnohi}[0]{\rwasm{}$_{ST}$\xspace}
\newcommand{\rwasmbaggy}[0]{\rwasm{}$_{S}$\xspace}
\newcommand{\rwasmnormalwasm}[0]{\rwasm{}$_{\wasm}$\xspace}
\newcommand{\graalvm}[0]{\text{GraalVM}\xspace}
\newcommand{\graalwasm}[0]{\text{GraalWasm}\xspace}
\newcommand{\graalwasmmswasm}[0]{\text{Graal}$_{ST}$\xspace}
\newcommand{\graalwasmnormalwasm}[0]{\text{Graal}$_{\wasm}$\xspace}

\newcommand{\evalrwasmbaggyovernormal}[0]{21.4\%\xspace}
\newcommand{\evalrwasmnohiovernormal}[0]{52.2\%\xspace}
\newcommand{\evalrwasmdefaultovernormal}[0]{197.5\%\xspace}
\newcommand{\evalgraalmswasmovernormal}[0]{42.3\%\xspace}
\newcommand{\evalrwasmnormalovernative}[0]{71.8\%\xspace}
\newcommand{\evalgraalnormalovernative}[0]{3230.0\%\xspace}
\newcommand{\locrwasm}[0]{1900\xspace}
\newcommand{\locgraal}[0]{1200\xspace}
\newcommand{\locllvm}[0]{1600\xspace}

\newcommand{\clang}[0]{\text{C}\xspace}

\makeatletter
\xdef\@thefnmark{\@empty}

\makeatother

\renewcommand{\emptyset}[0]{\varnothing}

\newcounter{hps}
\crefname{hps}{}{}

\newcommand{\proven}[1]{\ensuremath{\checkmark}}

\newcommand{\traceform}[1]{ { #1 }}

\newcommand{\rms}[1]{ \funt{RMS}{#1} }

\newcommand\xrsquigarrowfat[1]{%
\mathrel{%
\begin{tikzpicture}[baseline= {( $ (current bounding box.south) + (0,-0.5ex) $ )}]
  \node[inner sep=.7ex] (a) {$\scriptstyle #1$};
  \path[draw,implies-,double distance between line centers=1.5pt,decorate,
    decoration={zigzag,amplitude=0.7pt,segment length=1.2mm,pre=lineto,
    pre   length=4pt}]
    (a.south east) -- (a.south west);
\end{tikzpicture}}%
}
\newcommand{\xleadsto}[1]{%
   \if\relax\detokenize{#1}\relax
   \rightsquigarrow
   \else
   \mathrel{%
     \begin{tikzpicture}[%
       baseline={(current bounding box.south)}
       ]
       \node[%
       ,inner sep=.6ex
       ,align=center
       ] (tmp) {$\scriptstyle #1$};
       \path[%
       ,draw,<-
       ,decorate,decoration={%
         ,zigzag
         ,amplitude=0.7pt
         ,segment length=1.2mm,pre length=3.5pt
       }
       ]
       (tmp.south east) -- (tmp.south west);
     \end{tikzpicture}
   }
   \fi
 }
\newcommand{\rmsto}[1]{\xleadsto{#1}}
\newcommand{\rmstofat}[1]{\xrsquigarrowfat{#1}}

\newcommand{\absh}[0]{T}

\newcommand{\rmred}[4]{ #1 \vdash #2 \rmsto{#3} #4 }
\newcommand{\rmredfat}[4]{ #1 \vdash #2 \rmstofat{#3} #4 }

\newcommand{\st}[0]{\ensuremath{^*}}
\newcommand{\stk}[1]{\ensuremath{{#1}\st}}

\newcommand{\nil}[0]{\ensuremath{[]}}
\newcommand{\lookup}[2]{\ensuremath{#1[#2]}}

\newcommand{\memsaf}[1]{ \fun{MS}{#1} }

\newcommand{\semarr}[0]{\rightarrowtriangle}
\DeclareMathOperator\sem{\ensuremath{\semarr}}

\DeclareMathOperator\semt{\trg{\boldsymbol{\semarr}}}
\DeclareMathOperator\srcsemt{\src{\boldsymbol{\semarr}}}

\newcommand{\handletag}{\boxempty}
\newcommand{\datatag}{\medcirc}

\newcommand{\genhandletag}{}
\newcommand{\handlefull}{}
\newcommand{\handlesliced}{}

\newcommand{\sizeof}[1]{\bl{|}#1\bl{|}}

\newcommand{\libc}{\texttt{libc}\xspace}

\newcommand{\printnames}[1]{#1}

\newcommand{\diffgray}[1]{\colorbox{gray!30}{\ensuremath{#1}}}

\newcommand{\tid}[0]{n_{id}}

\newcounter{contrib}
\newcommand{\contribnum}[0]{\stepcounter{contrib}{\arabic{contrib}}.~}

\renewcommand{\OB}[1]{\stk{#1}}

\renewcommand{\genhandletag}{}
\renewcommand{\handlefull}{}
\renewcommand{\handlesliced}{}

\newcommand{\contribution}[1]{\smallskip\noindent\textbf{{#1.}\xspace}}

\def\and{\par}\normalbaselines

\DeclareUnicodeCharacter{03BB}{$\lambda$}

\AtBeginEnvironment{tikzpicture}{\catcode`$=3 } 

\colorlet{NAVYBLUE}{NavyBlue}

\setcopyright{none}

\acmConference{}

\begin{document}

\title[\mswasm]{\papertitle}         


\author{Alexandra E. Michael}
\orcid{}             
\affiliation{
  \institution{UCSD}            
  \country{USA}
}

\author{Anitha Gollamudi}
\orcid{}             
\affiliation{
  \institution{Yale University}            
  \country{USA}
}

\author{Jay Bosamiya}
\orcid{}             
\affiliation{
  \institution{CMU}            
  \country{USA}
}

\author{Craig Disselkoen}
\orcid{}             
\affiliation{
  \institution{UCSD}            
  \country{USA}
}

\author{Aidan Denlinger}
\orcid{}             
\affiliation{
  \institution{UCSD}            
  \country{USA}
}

\author{Conrad Watt}
\orcid{}             
\affiliation{
  \institution{University of Cambridge}            
  \country{UK}
}

\author{Bryan Parno}
\orcid{}             
\affiliation{
  \institution{CMU}            
  \country{USA}
}

\author{Marco Patrignani}
\orcid{0000-0003-3411-9678}
\affiliation{
  \institution{University of Trento}           
  \country{Italy}
}

\author{Marco Vassena}
\orcid{}             
\affiliation{
  \institution{Utrecht University}           
  \country{Netherlands}
}       

\author{Deian Stefan}
\orcid{}             
\affiliation{
  \institution{UCSD}            
  \country{USA}
}

\author{} 

\begin{abstract}
Most programs compiled to WebAssembly (\wasm) today are written in unsafe
languages like C and C++.
Unfortunately, memory-unsafe C code remains unsafe when compiled to \wasm---and
attackers can exploit buffer overflows and use-after-frees in \wasm almost as
easily as they can on native platforms.
Memory-Safe WebAssembly (\mswasm) proposes to extend \wasm with language-level
memory-safety abstractions to precisely address this problem.
In this paper, we build on the original \mswasm position paper to realize this
vision.
We give a precise and formal semantics of \mswasm, and prove that
well-typed \mswasm programs are, by construction, robustly memory
safe.
To this end, we develop a novel, language-independent memory-safety
property based on \emph{colored} memory locations and pointers.
This property also lets us reason about the security
guarantees of a formal \clang-to-\mswasm compiler---and prove
that it always produces memory-safe programs (and preserves the
semantics of safe programs).
We use these formal results to then guide several implementations: Two
compilers of \mswasm to native code, and a \clang-to-\mswasm compiler (that
extends Clang).
Our \mswasm compilers support different enforcement mechanisms, allowing
developers to make security-performance trade-offs according to their needs.
Our evaluation shows that the overhead of enforcing memory safety in software
ranges from 22\% (enforcing spatial safety alone) to 198\% (enforcing full
memory safety) on the PolyBenchC suite.
More importantly, \mswasm's design makes it easy to swap between enforcement
mechanisms; as fast (especially hardware-based) enforcement techniques
become available, \mswasm will be able to take advantage of these
advances almost for free.

\begin{center}\small\it
	{In the following, we use syntax highlighting accessible to both colourblind and black \& white readers \citep{patrignani2020use}. Specifically, we use a \src{blue},
	\src{sans\text{-}serif} font for \src{\clang} and a \trg{red},
	\trg{bold} font for \trg{\mswasm}.
	}
\end{center}
\end{abstract}



\maketitle


\section{Introduction}\label{sec:intro}

WebAssembly (\wasm) is a new bytecode designed to run native
applications---e.g., applications written in C/C++ and Rust---at native speeds,
everywhere---from the Web, to edge clouds, and IoT platforms.
%
Unlike most industrial bytecode and compiler intermediate
representations, \wasm was designed with safety in mind: \wasm
programs run in an isolated sandbox \emph{by construction}.
On the Web, this means that \wasm programs cannot read or corrupt the
browser's memory~\cite{Haas:2017}.
On edge clouds, where \wasm programs written by different clients run
in a single process, this means that one client cannot interfere with
another~\cite{lucet-talk}.

Within the sandbox, however, \wasm offers little protection.
Programs written in unsafe languages---and two thirds of existing \wasm
programs are compiled from C/C++~\cite{Lehmann21}---remain unsafe when
compiled to \wasm~\cite{Lehmann20}.
Indeed, buffer overflows and use-after-free vulnerabilities are as easy to exploit in \wasm as
they are natively; sometimes even \emph{easier} (e.g., because \wasm lacks
abstractions like read-only memory).
Worse, attackers can use such exploits to confuse the code hosting
\wasm into perfoming unsafe actions---to effectively bypass the \wasm sandbox.
\citep{Lehmann20}, for example, show how attackers can turn a buffer overflow
vulnerability in the \texttt{libpng} image processing library (executing in a
\wasm sandbox) into a cross-site scripting (XSS) attack.

To prevent such attacks, C/C++ compilers would have to insert
memory-safety checks \emph{before} compiling to \wasm---e.g., to
ensure that pointers are valid, within bounds, and point to memory
that has not been freed~\cite{softbound,cets,ccured}.
Industrial compilers like Emscripten and Clang do not.
Also, they \emph{should not}.
%
Retrofitting programs to enforce memory safety gives up on \emph{robustness},
i.e., preserving memory safety when linking a (retrofitted) memory-safe module
with a potentially memory-unsafe module.
It gives up on \emph{performance}: efficient memory-safety enforcement
techniques rely on operating system abstractions (e.g., virtual
memory~\cite{oscar}), abuse platform-specific details (e.g., encoding bounds
information in the (unused) upper bits of an address~\cite{baggy-bounds}), and
take advantage of hardware extensions (e.g., Arm's pointer authentication and
memory tagging extensions~\cite{armmte,pac}).
Finally, it also makes it harder to prove that memory safety is preserved
end-to-end.
%

With \emph{Memory-Safe WebAssembly}, \citep{mswasm} propose to bridge this gap
by extending \wasm with language-level memory-safety abstractions.
In particular, \mswasm extends \wasm with \emph{segments}, i.e., linear regions of
memory that can only be accessed using \emph{handles}.
Handles, like CHERI capabilities~\cite{cheri}, are unforgeable, well-typed
pointers---they encapsulate information that make it possible for \mswasm
compilers to ensure that each memory access is valid and within the segment
bounds.
Alas, the \mswasm position paper only outlines this design---they do not give
a precise semantics for \mswasm, nor implement or evaluate \mswasm as a
memory-safe intermediate representation.

This paper builds on this work to realize the vision of \mswasm.
We do this via five contributions:

\contribution{\contribnum Semantics and Memory Safety for \mswasm (\Cref{sec:mswasm-lang})}
Our first contribution is a formal specification of \mswasm as an extension of the
\wasm language, type system, and operational semantics.
Our semantics give precise meaning to the previous informal design~\cite{mswasm}.
Moreover, these semantics allow us to prove that all well-typed \mswasm programs are
\emph{robustly memory safe}; i.e., \mswasm programs are memory safe when linked against
arbitrary code.


\contribution{\contribnum Color-based Memory-Safety Monitor (\Cref{sec:mswa-prop})}
We develop a novel, abstract memory-safety monitor based on
\emph{colored} memory locations and pointers, which we use to show
that \mswasm is memory safe.
Colors abstract away specific mechanisms that \mswasm backends can
employ to enforce memory safety.  Additionally, they enable reasoning
about spatial as well as temporal memory safety, both at the
granularity of individual memory objects and within structured objects
as well.
Furthermore, since our memory-safety monitor is language-independent,
we can reason about memory-safety across compilation and establish the
soundness of our compiler-based memory-safety enforcement in our next
contribution.

%
%

\contribution{\contribnum Sound Compilation from C to \mswasm~(\Cref{sec:compiler})}
Like \wasm, \mswasm is intended to be used as a compilation target from
higher-level languages.
Hence, our third contribution is a formal \clang-to-\mswasm compiler,
which guarantees memory-safe execution of unsafe code.
In particular, we formalize a compiler from a subset of \clang to
\mswasm and prove that the compiler soundly \emph{enforces}
memory-safety.
Intuitively, this result ensures that memory-safe C programs when
compiled to \mswasm remain safe and preserve their semantics, while
memory-unsafe C programs trap at the first memory violation (and are
thus safe too).

\contribution{\contribnum Implementations of \mswasm (\Cref{sec:impl})}
Our next contribution is the implementation of three \mswasm-related compilers.
First, we implement an ahead-of-time (AOT) \mswasm-to-machine code compiler by
extending the rWasm~\cite{rwasm} compiler with \locrwasm lines of code (LOC).
%
Our extension of rWasm supports multiple options for enforcing
memory safety, with tradeoffs between performance and differing levels
of memory safety (spatial and temporal safety, and handle integrity).
Our second compiler is a just-in-time (JIT) \mswasm-to-JVM compiler (\locgraal LOC),
which uses the GraalVM Truffle framework~\cite{graalvm-truffle}.
Finally, our third compiler is an LLVM-to-\mswasm compiler (\locllvm LOC) created as an extension of the
CHERI Clang compiler toolchain~\cite{cheri-llvm}.

\contribution{\contribnum Evaluation of \mswasm (\Cref{sec:eval})}
Our final contribution is an empirical evaluation of \mswasm.
We benchmark \mswasm on PolyBenchC, the de-facto \wasm benchmarking
suite~\cite{polybench-c}.
We find that, on (geomean) average, \mswasm when enforced in software using our
AOT compiler imposes an overhead of \evalrwasmdefaultovernormal, which is
comparable with prior work on enforcing memory safety for C~\cite{cets}.
\mswasm, however, makes it easy to change the underlying
enforcement mechanism (e.g., to boost performance), without changing the
application.
To this end, we find that enforcing just spatial and temporal safety imposes
a \evalrwasmnohiovernormal overhead, and enforcing spatial safety alone using a
technique similar to Baggy Bounds~\cite{baggy-bounds}, is even
cheaper---\evalrwasmbaggyovernormal.
Our JIT compiler, which enforces spatial and temporal safety, but not
handle integrity, has an overhead of \evalgraalmswasmovernormal.
While these overheads are relatively large on today's hardware, upcoming
hardware features explicitly designed for memory-safety enforcement can reduce
these overheads (e.g., Arm's PAC can be used to reduce pointer integrity
enforcement to under 20\%~\cite{pac}, while Arm's CHERI~\cite{arm-cheri} or
Intel's CCC~\cite{intel-ccc} can also reduce the cost of enforcing temporal and
spatial safety).
\mswasm will be able to take advantage of these features as soon they
become available,
as illustrated by the ease of swapping memory-safety enforcement techniques
within our AOT compiler.


\contribution{Open Source \& Technical Report} Our technical report, implementations, benchmarks, and data sets are
available as supplementary material and will be made open source.

\section{Background and Motivation}\label{sec:bg}
We now give a brief introduction to \wasm (\Cref{sec:wa}), its attacker model (\Cref{sec:atk-model}), and the implications of memory unsafety within the \wasm sandbox
(\Cref{sec:unsafe-sandboxing}).
Then we give a brief introduction to \mswasm and to the open challenges we address in this work (\Cref{sec:mswasm-design}).
%


%
%
%
\subsection{WebAssembly}\label{sec:wa}
\wasm is a low-level bytecode, designed as a
safe compilation target for higher-level languages like C/C++ and Rust~\cite{wasm}.
\wasm bytecode is executed in a sandboxed environment by a stack-based
virtual machine.
Prior to execution, the virtual machine type-checks the bytecode to
ensure that each instruction finds the appropriate operands on the
stack.
\wasm's type system is extremely simple; the language has four primitive
types---32- and 64-bit integers and floats (\trg{i32} and \trg{i64}, and
\trg{f32} and \trg{f64} respectively)---and only structured control flow
constructs (i.e., no gotos) which simplify type checking.
The \wasm heap (or \textit{linear memory}), however, is an untyped contiguous
linear array of bytes.
Instructions \trg{\trgb{\tau}.load} and \trg{\trgb{\tau}.store} allow values of
the four primitive types to be read from and written to the memory at arbitrary
integer offsets.
At runtime, \wasm ensures that these accesses are in bounds (and \emph{traps}
when they are not).

This simple design makes whole classes of attacks impossible by design. For
example, the type-system ensures that well-typed bytecode cannot hijack the
virtual machine's control flow via stack-smashing
attacks~\cite{alephone1996smashing}. The coarse-grained bounds-checks on memory
accesses, together with structured control flow, confine \wasm to a
sandbox~\cite{wahbe93,tan:sfi}---and thus prevent \wasm from harming its host
environment.

This simple design has a trade-off: We necessarily lose information when
compiling programs written in high-level languages to \wasm.
Clang, for example, compiles complex source-level values (e.g., structs and
arrays) into ``bags of bytes'' in the untyped linear memory and compiles
pointers to \wasm 32-bit integers, offsets in the linear memory where
values are layed out.
This, unfortunately, means that misusing C/C++ pointers is as simple and severe
in \wasm as it is for native platforms.

\subsection{Threat Model}
\label{sec:atk-model}
In this work, we consider a \wasm-level attacker who attempts to exploit a
memory vulnerability present in a C program compiled to \wasm.
We consider vulnerabilities that can be triggered by \emph{spatial} memory
errors (e.g., buffer overflows), \emph{temporal} memory errors (e.g.,
use-after-free and double-free vulnerabilities), and \emph{pointer integrity}
violations (e.g., corrupting function pointers to bend control flow).
We assume the vulnerable program is linked with arbitrary code
written by the attacker, which can interact with the program in any
way allowed by \wasm semantics.
To exploit a vulnerability, the attacker code can supply malicious inputs to
the program and abuse values (including pointers) returned by or passed to the
program.
We leave memory unsafety of C++ programs and type confusion vulnerabilities~\cite{typesan} for future work.

\subsection{Sandboxing Without Memory Safety}\label{sec:unsafe-sandboxing}
Memory unsafe C programs, when compiled to \wasm, largely remain unsafe:
They can run uninterrupted as long as their reads and writes stay within the
bounds of the entire linear memory.
Unfortunately, \wasm also lacks most mitigations we rely on today to deal with
memory unsafety (e.g., memory protection bits and ASLR), so a program compiled
to run within \wasm's sandbox may be more vulnerable than if it were running on
bare metal~\cite{Lehmann20}.
\begin{lstlisting}[language=C,label=lis:wa,caption={Vulnerable code adapted from libpng 1.6.37}, float=ht, escapechar=|]
char * trim_token( char * token ){
  char * trimmed = malloc( 1024 * sizeof(char) );
  int i = 0, j = 0;
  while (token[i++] == ' ');            |\label{line:c-while}|
  char next = token[--i];       
  while (next != '\0') {        |\label{line:c-while2}|        
    trimmed[j++] = next;  // Possible buffer overflow     |\label{line:over}|
    next = token[++i];
  } |\label{line:c-while3}|                  
  trimmed[i] = '\0';
  return trimmed;
}
\end{lstlisting}

To understand how source-level memory vulnerabilities persist across
compilation, consider the C code snippet in \Cref{lis:wa} from \texttt{libpng
1.6.37}.
Function \texttt{\color{blue}{trim\_token}} takes a
pointer to a null-terminated string as input and returns a pointer to
a dynamically-allocated copy of the string, trimmed of the leading
whitespace characters.
The first loop (\cref{line:c-while}) simply scans the string \texttt{\color{blue}{token}}
and skips all the whitespace characters, while the second loop (\cref{line:c-while2}--\cref{line:c-while3})
copies the rest of the string into \texttt{\color{blue}{trimmed}} one character
at the time, until it finds the null terminator.
The vulnerability is on \cref{line:over}: the length of
the string \texttt{\color{blue}{token}} after trimming may exceed the
size allocated for buffer \texttt{\color{blue}{trimmed}}.
To exploit this vulnerability, an attacker only needs to call this
function on a sufficiently long string (longer than 1024 characters
after trimming).
This will cause the function to write past the bounds of
\texttt{\color{blue}{trimmed}}, thus corrupting the memory of the
program with the payload supplied by the attacker.
%
This vulnerability remains in the code obtained by compiling function
\texttt{\color{blue}{trim\_token}} with existing \wasm compilers
(e.g., Emscripten and Clang).
%
In particular, \cref{line:over} gets translated into the \wasm
instructions in \Cref{lis:wasm-generated1}.

\begin{wrapfigure}[10]{R}{.25\textwidth}
\begin{minipage}{0.23\textwidth}
\begin{lstlisting}[escapechar=|,label=lis:wasm-generated1,caption="Compilation of \Cref{line:over} into \wasm."]
|\trg{get\ \$trimmed}|
|\trg{get\ \$j}|
|\trg{i32.add}|
|\trg{get\ \$next}|
|\trg{i32.store} |
|\trg{...}  \quad \mbox{\emph{;; increment} $\trg{\$j}$ }|
\end{lstlisting}
\end{minipage}
\end{wrapfigure}
%
The first three \wasm instructions compute the address (a
32-bit integer) where the next character gets copied, by adding index
$\trg{\$j}$ to address $\trg{\$trimmed}$.
Then, instruction $\trg{get\ \$next}$ pushes the value of
the next character on the stack and $\trg{i32.store}$ writes it
to the address computed before.
As long as this address is within the linear memory region, the store
instruction succeeds---even if the address does not belong to the
buffer allocated for $\trg{\$trimmed}$.

Although an attacker could not use this memory-safety vulnerability to escape
\wasm's sandbox, they could use it to corrupt and steal data (e.g., private
keys) sensitive to the \wasm program itself.
\wasm programs on the Web already handle sensitive data,
and as \wasm's adoption expands beyond the Web,
addressing memory safety within the sandbox is crucial.

\subsection{The \mswasm Proposal}\label{sec:mswasm-design}

Memory-Safe WebAssembly (\mswasm) addresses these challenges by extending \wasm
with abstractions for enforcing memory safety~\cite{mswasm}.
Specifically, \mswasm introduces a new memory region called \emph{segment
memory}.
The segment memory consists of individual \emph{segments}, which are linearly
addressable, bounded regions of memory representing dynamic memory
allocations.
Unlike \wasm's linear memory, the segment memory cannot be accessed at
arbitrary offsets through standard \trg{load} and \trg{store} instructions.
Instead, \mswasm provides new types, values, and instructions to
regulate access to segments and enforce per-allocation memory safety.
Segments can only be accessed through \emph{handles}, unforgeable
memory capabilities that model pointers bounded to a particular
allocation of the segment memory.
\mswasm adopts this low-level memory model since an object-based model
(like that of the JVM) would be an inefficient (due to garbage
collection overhead) and overly restrictive (due to the constraints
of an object-based type-system) compilation target for C code
deployed to \wasm.

%
%
%
Handles are tuples: \trg{\langle base, offset, bound, isCorrupted, id
  \rangle}, where \trg{base} represents
the beginning of the segment in segment memory,
\trg{offset} is the handle's offset \emph{within} the segment,
i.e., within the \trg{bound}, that the handle points to.
Thus, a handle points to the address given by \trg{base + offset}.
%
%
%
%
\mswasm guarantees \emph{handle integrity} using
the \trg{isCorrupted} flag.
Intuitively, attempts to forge handles (e.g., by casting an integer,
or altering the bitstring representation of an existing handle in
memory) result in a corrupted handle.
\mswasm traps only when an out-of-bounds or corrupt handle is \emph{used}, not
when it is created.
This improves both performance, by eliminating checks on every
pointer-arithmetic operation, and compatibility, since
many C idioms create benign out-of-bound
pointers~\cite{checkedc,pointer-provenance,into-the-depths-of-C}.
Finally, \mswasm associates each segment allocation with a unique
identifier $\trg{id}$, which is used to enforce \emph{temporal}
memory safety.
%

%
%
%
\begin{wrapfigure}[13]{R}{.22\textwidth}
\begin{minipage}{0.2\textwidth}
%
\begin{lstlisting}[escapechar=|,label=lis:wasm-generated2,caption="Compilation of \Cref{lis:wa} into \wasm."]
|\trg{i32.const\ 1024}|
|\trg{\newst} |
|\trg{set\ \$trimmed} |
|\trg{...} |
|\trg{get\ \$trimmed} |
|\trg{get\ \$j} |
|\trg{\add} |
|\trg{get\ \$next} |
|\trg{\hstoretau{i32}}|
|\trg{...}|
\end{lstlisting}
\end{minipage}
\end{wrapfigure}
\mswasm provides new instructions to create and manipulate handles,
and to access segments safely through them.
Instructions \hload\ and \hstore\ are analogous to
\trg{\trgb{\tau}.load} and \trg{\trgb{\tau}.store}, but operate on
handles and trap if the handle is corrupted or points outside the segment bounds,
or if the segment has been freed.
Instruction \newst\ allocates a segment of the desired size in a free
region of segment memory and returns a handle to it.
Instruction \freest\ frees the segment associated with a valid handle,
thus making that region of segment memory available for new
allocations.
Lastly, instruction \add\ is for pointer arithmetic and modifies the
handle offset, without changing the base or bound.

%
%

%
%


%
\myparagraph{Eliminating unsafety by compiling C to \mswasm.}
With \mswasm we can eliminate potential memory
vulnerabilities automatically, via compilation.
For example, a C to \mswasm compiler would emit the instructions in \Cref{lis:wasm-generated2} for
the code snippet from~\Cref{lis:wa}.
This code allocates a new $\trg{1024}$-byte segment and stores the handle for
it in variable \trg{\$trimmed}.
Then, the \add\ instruction increments the offset of \trg{\$trimmed} with
index $\trg{\$j}$ and instruction \trg{\hstoretau{i32}} writes
$\trg{\$next}$ in the segment.
Since \mswasm instructions enforce memory safety, this code is safe to
execute even with malicious inputs.
In particular, if the offset of \trg{\$trimmed} is incremented past
the bound of the handle, the store instruction simply traps, thus
preventing the buffer overflow.

%
%



\myparagraph{Enforcing Intra-Object Memory Safety.}
Through the abstractions described above, \mswasm enforces
\emph{inter-object} memory safety, i.e., at the
granularity of individual allocations.
%
Unfortunately, this alone is insufficient to prevent memory-safety
violations within composite data types (e.g., structs), in which a
pointer to a field overflows (or overruns) an adjacent field.

\begin{lstlisting}[language=C,label=lis:struct, caption={Intra-object memory safety vulnerability.}, float=ht]
struct User {char name[32], int id };
struct User * my_user = malloc(sizeof(struct User));
char * my_name = my_user->name;
...
\end{lstlisting}
Consider the code snippet in \Cref{lis:struct}, which defines a
struct object containing a fixed-length string
\texttt{\color{blue}{name}} and an integer user \texttt{\color{blue}{id}}.
When compiled to \mswasm, this code allocates a single segment for the
\texttt{\color{blue}{User}} structure; thus the handle corresponding
to \texttt{\color{blue}{my\_name}} and derived from
\texttt{\color{blue}{my\_user}} via pointer arithmetic can also access
field \texttt{\color{blue}{id}} without trapping.
Therefore, an attacker could exploit a memory vulnerability in the
code that manipulates \texttt{\color{blue}{my\_name}} to corrupt the
user \texttt{\color{blue}{id}} and impersonate another user.
%

Hence, to enforce \emph{intra-object} memory safety, \mswasm provides
an additional instruction called $\hslice$.
Instruction \hslice\ shrinks the portion of the segment that a handle
can access by growing its $\trg{base}$ and reducing its $\trg{bound}$
field by a given offset.
By emitting a \hslice\ instruction with appropriate offsets for
expression \texttt{\color{blue}{my\_user->name}}, a compiler can
generate a sliced handle that includes only the
\texttt{\color{blue}{name}} field.
As a result, if the attacker later tries to overflow
\texttt{\color{blue}{my\_name}}, the safety checks of the sliced
handle will detect a violation and trap the execution, thus preventing
the program from corrupting the user \texttt{\color{blue}{id}}.



\myparagraph{The missing pieces.}
The original \mswasm position paper~\cite{mswasm} only outlines the basic
abstractions we describe above.
The position paper does not give a formal (or even informal) semantics for the
proposed language extensions.
They do not describe compilation techniques---how one would compile C code to
\mswasm or how \mswasm would be compiled to native code---nor an implementation
(and thus evaluation) of \mswasm.
In this paper we address these limitations and, for the first time,
provide an end-to-end, robust, memory-safe \clang-to-\mswasm compiler
that is rooted in formal methods.



%

\section{The \mswasm Language}\label{sec:mswasm-lang}
%
%
This section develops a formal model of the design of \mswasm
described above.
The model includes syntax, typing (\Cref{sec:syn}), and operational semantics for \mswasm (\Cref{sec:sem}) and it serves as a specification for different low-level mechanisms (bounds checks, segment identifiers, integrity tags, etc.) needed to enforce memory safety in \mswasm.
We present the properties of \mswasm in the next section, after formally defining memory safety.


Due to space constraints, we present a selection of the
formalization, and elide proofs and auxiliary lemmas.
The interested reader can find these omissions in the supplementary material.

  %

%
%
%

\subsection{\mswasm Syntax}\label{sec:syn}
\myfig{
\begin{gather*}
    \begin{aligned}
    	\printnames{\text{Modules } }
    	\trg{M} \bnfdef
       		&\
       		\trg{ \{ funcs\ \trgfuncs, imports\ \trgb{\rho}^* , heap\ n_H, \hl{\trg{segment\ n_S}} \}}
    \end{aligned}
    \\
    \begin{aligned}
       	\printnames{\text{Fun. Defs } }
       	\trg{\fdef}\bnfdef
       		&\
       		\trg{ \{ var\ \stk{\trgb{\tau}}, body\ \stk{i} \} : \trgb{\rho}}
       	&
        \printnames{\text{Instructions Types } }
        \trgb{\rho} \bnfdef
          &\
          \trg{\stk{\trgb{\tau}} \rightarrow \stk{\trgb{\tau}}}
    \end{aligned}
    \\
    \begin{aligned}
      \printnames{\text{Value Types } }
      \trgb{\tau} \bnfdef
          &\
          \trg{\ittw} \mid \trg{i64} \mid \trg{f32} \mid \trg{f64}
          \mid \hl{\handle}
    \end{aligned}
    \\
    \begin{aligned}
    \printnames{\text{Instructions } }
    \trg{i} \bnfdef
    	&\
    	\trg{nop} \mid \trg{trap} \mid \trg{\trgb{\tau}.const\ c} \mid \bopt \mid \trg{get\ n} \mid \trg{set\ n}
    \mid
    	\trg{\trgb{\tau}.load} \mid \trg{\trgb{\tau}.store} \mid \iftt{\trg{i\st}}{\trg{i\st}} \mid \trg{call\ n}
    \\
    \mid
        &\
        \trg{return} \mid \hl{\hload} \mid \hl{\hstore} \mid \hl{\trg{slice}}
    \mid
    	\hl{\newst} \mid \hl{\add} \mid \hl{\freest}
    \end{aligned}
\end{gather*}
}{syntax}{Syntax of \mswasm with extensions to \wasm highlighted.}
%
%
%
%
%
%
%
%
The syntax of \mswasm is defined in \Cref{fig:syntax}.
\mswasm programs are modules \trg{M}, which specify a list of function
definitions \trgfuncs, the type of imported functions
\trgb{\stk{\rho}}, and the size of the linear and segment memory
(\trg{n_H} and \trg{n_S} $\in\mb{N}$).\footnote{
  We use \com{\stk{e}} to denote a list of \com{e} elements, and
  \com{e^n} for a list of length \com{n}.
  We write \com{[e_0,e_1,\cdots]} for finite lists,
  \com{\nil} for the empty list, \com{{e}\stackcat\stk{e}} to add
  \com{e} in front of \stk{e}, and \com{\stk{e_1}\stackapp\stk{e_2}}
  to append \stk{e_2} to \stk{e_1}.  Notation
  \com{\lookup{\stk{e}}{i}} looks up the i-th element of \stk{e} and
  \com{\stk{e}[i\mapsto v]} replaces the i-th element of \com{\stk{e}}
  with v.}
Syntax \trg{\{ var\ \stk{\trgb{\tau}}, body\ \stk{i} \} : \trgb{\rho}}
defines a function with local variables of types \trgb{\stk{\tau}},
body \trg{\stk{i}}, and function type \trgb{\rho}.
%
%
%
Instructions \trg{i} manage the operand stack and are mostly standard.
%
%
Variables are referred to through numeric indices \trg{n}, which are
statically validated during type-checking.
For example, instructions \trg{get\ n} and \trg{set\ n} retrieve and
update the value of the \trg{n}-th local variable, respectively.
Function calls are similar, i.e., instruction \trg{call\ n} calls
the \trg{n}-th function (either defined or imported) in the scope of
the module.
We describe \mswasm's instructions on segments and handles below.

\myparagraph{Typing.}
The type system of \mswasm is a straightforward extension of \wasm's,
and it similarly guarantees type safety (i.e.,
well-typed modules satisfy progress and preservation).
Instructions are typed by the judgment \trg{\trgb{\Gamma}\vdash i :
\trgb{\tau}_1\st \to \trgb{\tau}_2\st}, where
\trgb{\tau_1\st} and \trgb{\tau_2\st} are the types of the values
that \trg{i} pops and pushes on the stack, respectively, and the
typing context \trgb{\Gamma} tracks the type of the variables and
functions in scope.
Compared to \wasm, the only restriction imposed by the type system of \mswasm
is that it prevents programs from forging handles by reading raw
bytes from the unmanaged linear memory, i.e., \trg{\trgb{\Gamma}\vdash
  \trgb{\tau}.load: {[\trgb{i32}] \to [\trgb{\tau}]}} iff
\trg{\trgb{\tau} \neq \handle}.

\subsection{\mswasm Operational Semantics}\label{sec:sem}
To reason about the memory-safety guarantees of \mswasm, we define a
small-step labeled operational semantics, which generates events for
memory-relevant operations such as segment allocations and accesses.

\subsubsection{Semantics of \wasm}\label{sec:sem-intra-pre}

\myfig{
\begin{gather*}
  \begin{aligned}
      \printnames{\text{Store } }
      \trgb{\Sigma} \bnfdef
        &\
        \trg{(H,}
          \hl{\trg{T}
            \trg{,A)}}
      &
      \printnames{\text{Heaps } }
      \trg{H} \bnfdef
        &\
        \trg{b\st}
      &
    \printnames{\text{Bytes }}
      \trg{b} \in
        &\
        \{0..2^8\text{-} 1\}
    \\
    \printnames{\text{Constants }}
      \trg{c} &
    &
    \printnames{\hl{\text{Segments}}}\
      \trg{T} \bnfdef
        &\
        \trg{\stk{(b,t)}}
      &
      \printnames{\hl{\text{Allocators}}}\
      \trg{A} &
    \\
    \printnames{\text{Local Frames } }
    \trg{F} \bnfdef
      &\
      \trg{(\trgb{\theta}, \stk{i}, \stk{v})}
    &
    \printnames{\text{Locals } }
    \trgb{\theta} \bnfdef
      &\
      \trg{\stk{(n\mapsto v)}}
    &
      \printnames{\hl{\text{Segment Tags}}}\
      \trg{t} \bnfdef
        &\
        \trgb{\datatag} \mid \trgb{\handletag}
  \end{aligned}
  \\
  \begin{aligned}
    \printnames{\text{Values } }
    \trg{v} \bnfdef
      &\
      \trg{c} \mid \hl{\trg{h}}
    &
    \printnames{\hl{\text{Handles}}}\
    \trg{h} \bnfdef
    &\
    {\trg{\conf{n_{base}, n_{offset}, n_{bound}, b_{valid}, \tid}_{{\genhandletag}}}}
    \\
  \end{aligned}
  \\
\begin{aligned}
  \printnames{\hl{\text{Memory Events}}}\
  \at \bnfdef
    &\
    \trgb{\epsilon} \mid \rdt{h} \mid \wrtt{h}
    \mid
      \trg{salloc(h)} \mid \trg{sfree(h)} \mid \trapt
\end{aligned}
\end{gather*}
}{rts}{\wasm and \hl{\mswasm} runtime structures.}

\Cref{fig:rts} defines the runtime structures used in the semantics
judgment.
%
%
%
A local configuration $\conf{\trgb{\Sigma},F}$ consists of the store
\trgb{\Sigma} and the stack frame \trg{F} of the function currently
executing.
In \wasm, the store \trgb{\Sigma} contains only the unmanaged linear memory
\trg{H}, which is a list of bytes \trg{\stk{b}} of fixed length.
%
%
%
%
%
%
%
%
%
%
%
%
%
%
%
The local stack frame \trg{F} maintains the environment \trgb{\theta}
for variable bindings (mapping from variable indices to
values), a list of instructions \trg{i\st} to be executed, and the
operand stack \trg{v\st} for the values produced (and consumed) by
those instructions.
%
Values include constants \trg{c} and integers \trg{n}.

The semantics judgment
$\wstof{\conf{\trgb{\Sigma},F}}{\trgb{\alpha}}{\conf{\trgb{\Sigma}',F'}}$
indicates that under function definitions $\trgfuncs$, local
configuration $\conf{\trgb{\Sigma},F}$ executes a single instruction
and steps to $\conf{\trgb{\Sigma}',F'}$, generating event $\trgb{\alpha}$ (explained below).
The semantics features also a separate judgment for function calls and returns,
which is standard and omitted.

\myfig{\centering
  \typerule{Stack-Top}{
    \wstof
      {\conf{\trgb{\Sigma}, (\trgb{\theta}, i, \stk{v})}}
      {\trgb{\alpha}}
      {\conf{\trgb{\Sigma'}, (\trgb{\theta'}, \stk{i'}, \stk{v'})}}
  }{
    \wstof
      {\conf{\trgb{\Sigma}, (\trgb{\theta}, i\stackcat\stk{i}, \stk{v}\stackapp\stk{v_b})}}
      {\trgb{\alpha}}
      {\conf{\trgb{\Sigma}, (\trgb{\theta}, \stk{i'}\stackapp\stk{i}, \stk{{v'}}\stackapp\stk{v_b})}}
  }{stk}
        \and
  \typerule{Get}{
    \trg{\lookup{\trgb{\theta}}{n}} = \trg{v}
  }{
    \wstof
      {\conf{\trgb{\Sigma}, (\trgb{\theta}, get\ n, [])}}
      {}
      {\conf{\trgb{\Sigma}, (\trgb{\theta}, [], \trg{[ v ]} )}}
  }{get}
        \and
  \typerule{If-T}{
    \trg{n} \neq \trg{0}
  }{
    \wstof
      {\conf{\trgb{\Sigma}, (\trgb{\theta}, \iftt{\trg{{i_1}\st}}{\trg{{i_2}\st}}, \trg{[n]})}}
      {}
      {\conf{\trgb{\Sigma}, (\trgb{\theta}, \stk{i_1}, [])}}
  }{if-t}
        \and
  \typerule{Load}{
    \trg{0 \leq n}
    &
    \trg{n + \sizeof{\trgb{\tau}} < \sizeof{H}}
    &
    \trg{b^{\sizeof{\trgb{\tau}}}} = \trg{  [\ \Sigma.H[n + j] \mid\ j \in \{0..\sizeof{\trgb{\tau}} -1 \}\ ]}
    &
    \trg{v} = \trg{\trgb{\tau}.unpack(b^{\sizeof{\trgb{\tau}}})}
  }{
    \wstof
      {\conf{\trgb{\Sigma}, (\trgb{\theta}, \loadt, \trg{[n]})}}
      {}
      {\conf{\trgb{\Sigma}, (\trgb{\theta}, [], \trg{[v]})}}
  }{load}
}{wasm-intra-rules}{
  Semantics of \wasm (excerpts).
}

\Cref{fig:wasm-intra-rules} presents a selection of rules that \mswasm inherits from \wasm.
Auxiliary \cref{tr:stk} extracts the first instruction and its
operands from the list of instructions and the stack, respectively,
and executes the instruction using the rules for individual
instructions.
\Cref{tr:get} executes instruction \trg{get\ n}, which
looks up the value of variable \trg{n} in the environment
\trgb{\theta}, i.e., $\trg{\lookup{\trgb{\theta}}{n}} = \trg{v}$, and
pushes \trg{v} on the stack.
Instruction $\trg{\iftt{\stk{i_1}}{\stk{i_2}}}$ pops the integer
condition \trg{n} from the stack and returns instructions
\trg{\stk{i_1}} if \trg{n} is non-zero via
\cref{tr:if-t}.\footnote{\wasm does not provide instructions for
  \emph{unstructured control-flow}, common on native architectures
  (e.g., \texttt{JMP} on x86). \wasm code can define and jump to typed
  \emph{labeled} blocks. Since these features do not affect the memory
  safety guarantees of \mswasm, we omit them from our model.}
\Cref{tr:load} loads a value of type \trgb{\tau} from address \trg{n}
in linear memory.
Since the linear memory consists of plain bytes, the rule reads
\sizeof{\trgb{\tau}} bytes at address \trg{n} into byte string
\trg{b^{\sizeof{\trgb{\tau}}}} and converts them into a value of type
\trgb{\tau}, i.e., \trg{v} =
\trg{\trgb{\tau}.unpack(b^{\sizeof{\trgb{\tau}}})}, which is then
pushed on the stack.
%
In the rule, premises \trg{0 \leq n} and \trg{n +
  \sizeof{\trgb{\tau}} < \sizeof{H}} ensure that the load instruction does
not read outside the bounds of the linear memory, but do not enforce
memory safety, as explained above.



\subsubsection{Semantics of \mswasm}\label{sec:sem-intra-mswa}


\mswasm extends the runtime structures of \wasm with a managed
segment memory, handle values, and a memory allocator (see
\Cref{fig:rts}).
The segment memory \trg{T} is a fixed-length list \trg{\stk{(b,t)}} of
\emph{tagged} bytes, where each tag \trg{t} indicates whether the
corresponding byte is part of a numeric value (\trg{t} =
\trg{\datatag}) or a handle (\trg{t} = \trgb{\handletag}).
These tags are used to detect forged or corrupted handles stored in
segment memory and thus ensure handle integrity.

%
Handles \trg{\conf{n_{base}, n_{offset}, n_{bound},
    b_{valid},\tid}_{{\genhandletag}}} contain the base address
\trg{n_{base}} of the memory region they span, length
\trg{n_{bound}}, offset \trg{n_{offset}} from the base,
integrity flag \trg{b_{valid}} which indicates whether the handle is
authentic (\trg{b_{valid}} = \trg{true}) or corrupted (\trg{b_{valid}}
= \trg{false}), and segment identifier \trg{\tid}.
%
%
%
%
%
%
%
%
%
%
Finally, \mswasm instructions generate memory events \trg{\alpha},
which include the silent event $\trgb{\epsilon}$, reading and writing
values of type \trgb{\tau} through a handle \trg{h} (i.e., $\rdt{h}$
and $\wrtt{h}$), segment allocations $\trg{salloc(h)}$, segment free
$\trg{sfree(h)}$, and $\trapt$ which is raised in response to a memory
violation.

\myparagraph{Memory Allocator.}
The \mswasm runtime system is responsible for providing a memory
allocator to serve memory allocations of compiled programs.
%
%
%
In our model, we represent the state of the memory allocator and its
semantics explicitly, as this simplifies reasoning about memory
safety.
The allocator state \trg{A} keeps track of free and used regions of
segment memory and their identifiers, i.e., \trg{A.free} and
\trg{A.allocated}, respectively.
The allocator serves allocation requests via reductions of the form
$\astep{\conf{T, A}}{\psalloct{a}{n,\tid}}{\conf{T', A'}}$, which
allocates and initializes a free segment of $\trg{n}$ bytes, which
starts at address \trg{a} in segment memory and can be identified by
fresh identifier \trg{\tid}.
Dually, reductions of the form
$\astep{\conf{T, A}}{ sfree(a,\tid)}{\conf{T', A'}}$ free the segment
identified by $\trg{\tid}$ and allocated at address \trg{a}, or
traps, if no such segment is currently allocated at that
address.
%
%
We omit further details about the allocator state and semantics---the
memory-safety guarantees of \mswasm do not depend on the concrete allocation
strategy.

\myfig{\centering
	\typerule{H-Load}{
		\trgb{\tau} \ne \trg{handle}
		&
		\trg{ v_1 } = \trg{ \conf{n_1, o, n_2, true,\tid} }
    &
    \trg{0 \leq o}
    &
    \trg{(\stk{b}, \_) } = \trg{  [\ \trgb{\Sigma}.T[n + j] \mid\ j \in \{0..\sizeof{\trgb{\tau}} -1 \}\ ]}
    \\
    \trg{o + \sizeof{\trgb{\tau}} < n_2}
    &
		\trg{n  } = \trg{  n_1 + o }
		&
		\trg{\tid\in \trgb{\Sigma}.A.allocated}
    &
    \trg{v_2  } = \trg{  \trgb{\tau}.unpack(\stk{b})}
	}{
		\wstof
			{\conf{\trgb{\Sigma}, (\trgb{\theta}, \hload , [ v_1 ] )}}
			{\readt{v_1}}
			{\conf{\trgb{\Sigma}, (\trgb{\theta}, \nil, [ v_2 ] )}}
	}{h-load}
	\and
	\typerule{H-Load-Handle}{
		\diffgray{
			\trgb{\tau} = \trg{\handle}
		}
		&
		\trg{ v_1  } = \trg{  \conf{n_1, o, n_2, true,\tid} }
    &
    \trg{0 \leq o}
    &
    \trg{ (\stk{b} , \stk{t})  } = \trg{  [\ \trgb{\Sigma}.T[n + j] \mid\ j \in \{0..\sizeof{\trgb{\tau}} -1 \}\ ] }
    \\
    \trg{o + \sizeof{\trgb{\tau}} < n_2}
    &
    \trg{ n  } = \trg{n_1 + o}
    &
    \trg{\tid\in \trgb{\Sigma}.A.allocated}
		&
		\diffgray{
			\trg{n\ \%\ \sizeof{\handle} } = \trg{0}
		}
		\\
		\diffgray{
			\trg{ b_c  } = \trg{  \bigwedge_{t \in \stk{t}}  (t = \trgb{\handletag}) }
		}
		&
		\trg{  \trgb{\tau}.unpack(\stk{b}) } = \trg{\conf{n_1',a',n_2',b_c',\tid'}  }
    &
		\trg{v_2} = \trg{   \conf{n_1',a',n_2', b_c \diffgray{ \trg{\wedge b_c'} },\tid' } }
	}{
		\wstof
			{\conf{\trgb{\Sigma}, (\trgb{\theta}, \hload  , [v_1] ) }}
			{\readt{v_1}}
			{\conf{\trgb{\Sigma}, (\trgb{\theta}, \nil, [v_2]) }}
	}{h-load-handle}
  \and
	\typerule{H-Store}{
    \trgb{\tau} \ne \trg{handle}
    &
    \trg{ v_1} = \trg{\conf{n_1, o, n_2, true,\tid} }
    &
    \trg{o \ge 0}
    &
    \trg{o + \sizeof{\trgb{\tau}} < n_2 }
    &
    \trg{\tid\in \trgb{\Sigma}.A.allocated}
    \\
    \trg{b^*} = \trg{\trgb{\tau}.pack(v_2)}
      &
    \trg{t} = \trg{\datatag}
    &
     \trg{ a = (n_1 + o)}
    &
    \trg{\trgb{\Sigma}'.T} = \trg{\trgb{\Sigma}.T[a + j \mapsto (b_j , t) \ \mid\ j \in \{0..\sizeof{\trgb{\tau}}-1\}]}
  }{
    \wstof
      {\conf{\trgb{\Sigma}, \trgb{\theta}, \hstore  , [v_2, v_1] }}
      {\writet{v1,v2}}
      {\conf{\trgb{\Sigma}', \trgb{\theta}, \nil, \nil}}
  }{h-store}
  \and
  \typerule{H-Store-Handle}{
    \diffgray{
      \trgb{\tau} = \trg{handle}
    }
    &
    \trg{v_1} = \trg{\conf{n_1, o, n_2, true, \tid} }
    &
    \trg{o \ge 0}
    &
    \trg{o + \sizeof{\trgb{\tau}} < n_2 }
    \\
    \trg{\tid\in \trgb{\Sigma}.A.allocated}
    &
    \trg{b^*} = \trg{\trgb{\tau}.pack(v_2)}
    &
    \diffgray{
      \trg{t} = \trgb{\handletag}
    }
    \\
      \trg{ a} = \trg{n_1 + o} 
    &
    \diffgray{
      \trg{a \% \sizeof{handle} } = \trg{0}
    }
    &
    \trg{ \trgb{\Sigma}'.T} = \trg{\trgb{\Sigma}.T[a + j \mapsto (b_j , t) \ \mid\ j \in \{0..\sizeof{\trgb{\tau}}-1\}]}
  }{
    \wstof
      {\conf{\trgb{\Sigma}, \trgb{\theta}, \hstore  , [v_2 , v_1] }}
      {\writet{v1,v2}}
      {\conf{\trgb{\Sigma}', \trgb{\theta}, \nil, \nil}}
  }{h-store-handle}
	\and
	\typerule{H-Alloc}{
	\trg{ \trgb{\Sigma}} =  \trg{(H, T, A)}
        &
	\astep
		{\conf{T , A}}
		{ \psalloct{a}{n,\tid} }
		{\conf{T' , A'}}
                &
		\trg{v  } = \trg{  \conf{a, 0, n, true,\tid}}
		&
		\trg{\trgb{\Sigma}' = (H, T', A')}
	}{
		\wstof
			{\conf{\trgb{\Sigma}, (\trgb{\theta}, \newst, [n])}}
			{salloc(v)}
			{\conf{\trgb{\Sigma}', (\trgb{\theta}, \nil,  [v])}}
	}{h-alloc}
  \typerule{H-Free}{
    \trgb{\Sigma} = \trg{(H, T, A)}
    &
    \trg{h} = \trg{\conf{a, 0, \_, true,\tid}_{\handlefull}}
    &
    \trg{
    \astep
      {\conf{T, A}}
      {{sfree(a,\tid)}}
      {\conf{T', A'}}
    }
    &
    \trg{\trgb{\Sigma}'} = \trg{(H, T', A')}
  }{
      \wstof
      {\conf{\trgb{\Sigma}, \trgb{\theta}, \freest , [h] }}
      {{sfree(h)}}
      {\conf{\trgb{\Sigma}', \trgb{\theta}, \nil,\nil}}
  }{h-free-full}
	\and
	\typerule{Handle-Add}{
		\trg{v  } = \trg{  \conf{n_1, o , n_2, b,\tid}}
		&
		\trg{v'  } = \trg{  \conf{n_1, o + n , n_2, b,\tid}}
	}
	{
		\wstof
			{\conf{\trgb{\Sigma}, (\trgb{\theta}, \add, [n , v])}}
			{}
			{\conf{\trgb{\Sigma}, (\trgb{\theta}, \nil, [ v' ])}}
	}
	{handle-add}
	\and
	\typerule{Slice}{
		\trg{v  } = \trg{  \conf{n_1, o , n_2, b,\tid}}
		&
		\trg{0} \leq \trg{o_1} < \trg{n_2}
		&
		\trg{0 } \leq \trg{o_2}
		&
		\trg{v'  } = \trg{  \conf{n_1 + o_1, o , n_2 - o_2, b,\tid}}
	}{
		\wstof
			{\conf{\trgb{\Sigma}, (\trgb{\theta}, \hslice, [v, o_2, o_1])}}
			{}
			{\conf{\trgb{\Sigma}, (\trgb{\theta}, \nil, [ v' ])}}
	}{slice}
}{mswasm-intra-rules}{
  Semantics of \mswasm (excerpts). The premises that ensure handle integrity are highlighted.
}

\myparagraph{\mswasm Rules.}
\Cref{fig:mswasm-intra-rules} gives some important rules for the new
instructions of \mswasm.
\Cref{tr:h-load} loads a non-handle value
($\trgb{\tau} \ne \trg{handle}$) from segment memory through a
\emph{valid} handle \trg{ \conf{n_1, o, n_2, true, \tid}_{\genhandletag}}.
%
\Cref{tr:h-load} reads bytes \trg{\stk{b}}
from the address pointed to by the handle, i.e.,
$\trg{n} = \trg{n_1 + o}$, and converts them into a value of type
$\trgb{\tau}$, i.e.,
$\trg{v_2 } = \trg{ \trgb{\tau}.unpack(\stk{b})}$.%
\footnote{Total function \trg{\trgb{\tau}.unpack} converts
  \sizeof{\trgb{\tau}} bytes (the number of bytes needed to
  represent a value of type \trgb{\tau}) into a value of type
  \trgb{\tau}. The inverse function \trg{\trgb{\tau}.pack}
  converts values to their byte representation.}
The rule enforces memory safety by checking that (1) the handle is
\emph{not} corrupted, (2) the load does not read bytes outside the
bounds of the segment, i.e., \trg{0 \leq o} and \trg{o +
  \sizeof{\trgb{\tau}} < n_2}, and (3) the segment is still allocated,
i.e., \trg{\tid\in A.allocated}.
%
%
%
\Cref{tr:h-load-handle} is similar, but for loading values of type
\trg{\handle}; therefore it includes additional checks (highlighted in
gray), to enforce handle integrity.
%
%
First, the rule checks that all the bytes read from memory are tagged
as handle bytes, i.e.,
$\trg{b_c} = \trg{\bigwedge_{t \in \stk{t}} (t = \trgb{\handletag})}$,
and then combines this flag with the flag \trg{b_c'} obtained from the
raw bytes of the segment; i.e., it returns handle
$\trg{\conf{n_1',a',n_2',b_c \wedge b_c',\tid }_{\genhandletag}}$.
The combined flag invalidates handles obtained from bytes tagged as
data, thus preventing programs from forging handles by altering their
byte representation in memory.
%
Furthermore, to enforce handle integrity, the rule allows loading
handle values only from \trg{\sizeof{handle}}-aligned memory
addresses, i.e.,
$\trg{ (n_1 + o)\% \sizeof{\handle} } = \trg{0}$.
The alignment requirement is needed to avoid crafting fake handles.
In fact, if one were to store two handles next to each other and then load from an address \emph{within} the first one, the load would succeed and load bytes that all have the capability tag. 
However, the loaded value would be a fake capability, since the loaded bytes would be part of the first capability, and part of the second.
Loading and storing at aligned addresses prevents this issue.
%
%
%
%
%
The rules for \hstore\ are analogous---they include similar bounds
checks and alignment restrictions for handles---and additionally set
the tag of the bytes that they write in memory according to
\trgb{\tau}.
For example, \Cref{tr:h-store} applies to values whose type are not
handle, therefore it tags the bytes of the value written to memory as
data (\trg{\datatag}).
In contrast, \Cref{tr:h-store-handle} writes a \trg{handle} to memory
and so it tags its bytes accordingly (i.e., \trg{\handletag}).
%

\Cref{tr:h-alloc} invokes the allocator to allocate and initialize a
new segment of $\trg{n}$ bytes at address \trg{a} in segment memory,
and returns a handle to it.
\Cref{tr:h-free-full} invokes the allocator to free the segment bound
to the given \emph{valid} handle.
%
%
%
%
\Cref{tr:handle-add} increments the offset of a handle \trg{v},
without changing the other fields.
Notice that this rule allows programs to create handles that point out of
bounds; out-of-bounds handles only cause a trap when they are used to access
memory.
\Cref{tr:slice} creates a sliced handle
$\trg{ \conf{n_1 + o_1, o , n_2 - o_2, b, \tid}_{\handlesliced} }$, where
the base is increased by offset \trg{o_1} and the bound is reduced by
offset \trg{o_2}.
Premises $\trg{0} \leq \trg{o_1} < \trg{n_2}$ and
$\trg{0 } \leq \trg{o_2}$ ensure that the handle obtained after
slicing can only access a subset of the segment accessible from the
original handle.
%

Whenever a \hload, a \hstore, a \freest, or a \hslice\ do not match their premise, the semantics traps, emitting a \trapt action and halting the execution immediately, with no values on the operand stack (omitted for brevity).

\section{Abstract Memory-Safety Monitor}\label{sec:mswa-prop}

This section presents an abstract notion of memory safety that is
%
based on \emph{colored} memory locations (\Cref{sec:ms}).
Colors soundly abstract away many implementation details,
which in turn let us formalize memory safety compactly as a trace
property checked by a corresponding monitor
(\Cref{subsec:ms-property}).
We use this monitor to establish the spatial and
temporal memory-safety guarantees of \mswasm (\Cref{sec:ms-in-mswa}).
Since our monitor is language-independent, we will reuse it to prove our \clang-to-\mswasm secure
compiler enforces memory safety~(\Cref{sec:compiler}).

%

\subsection{Color-based Memory Safety}\label{sec:ms}
%


Our notion of memory safety associates pointers and memory locations
with \emph{colors} (which represent pointer provenance~\cite{c-provenance}), shades, and allocation tags.
Intuitively, each memory allocation generates a pointer annotated with
a \emph{unique} color (and shades as described below) and assigns the same color to each location in
the allocated region of memory.
Then, we consider a memory access \emph{spatially} safe if the color
of the pointer corresponds to the color of the memory location it
points to.
To account for \emph{temporal} safety, memory locations are tagged as free or allocated
and we enforce that accessed locations are tagged as allocated.
Colors are suitable to reason about memory safety at the granularity
of individual memory objects.
In particular, this simple model is sufficient for low-level languages
that do not natively support composite data types (e.g., \wasm and
\mswasm).
However, colors alone cannot capture \emph{intra-object} memory
violations (e.g., the vulnerability in \Cref{lis:struct}).
Intuitively, this is because the simple model assigns the \emph{same}
color to all the fields of a struct object.
To reason about intra-object safety, we thus extend colors with
\emph{shades} and use a different shade to decorate the memory
locations of each field in a struct.
As a result, a pointer to a struct field cannot be used to access
another field of the same struct, as their shades do not match.

As explained above, our definition of memory safety is intentionally
minimal and language-agnostic: it does not specify other operations on
colored pointers, e.g., pointer arithmetic, and how they propagate
colors.
This lets us reuse this definition of memory safety for different
languages and reason about enforcing memory-safety via compilation
in~\Cref{sec:compiler}.


%


\subsection{Memory-Safety Monitor}\label{subsec:ms-property}
We formalize our notion of memory safety by constructing a safety
monitor~\cite{Schneider20}, i.e., a state machine that checks whether
a trace satisfies memory safety.
Intuitively, the monitor consumes a trace of memory events and gets stuck when it encounters a
memory violation.
%
We assume an infinite set of colors $\com{\mc{C}}$, shades
$\com{\mc{S}}$, and define a \emph{colored shadow memory}
$\com{\absh} \in \mathbb{N} \rightharpoonup \{A(c,s) , F(c,s) \}$,
i.e., a finite partial map from addresses $\com{a \in \mathbb{N}}$ to
tagged colors $\com{c} \in \com{\mc{C}}$ and shades
$\com{s} \in \com{\mc{S}}$, where tags $A$ and $F$ denote whether a
memory location is allocated or free, respectively.
Then, we define an \emph{abstract trace model} of memory events
$\com{\ac}$, which include read and write operations with colored
pointers, i.e., $\com{\rd{a}{(c,s)}}$ and $\com{\wrt{a}{(c,s)}}$,
memory allocations, i.e., $\com{\sallocgen{n}{a}{c}{\phi}}$ denoting a
\com{n}-sized \com{c}-colored allocation starting at address
$\com{a}$, in which sub-regions are shaded according to function
$\com{\phi} : \{ 0,\ldots,n-1 \} \rightarrow \mc{S}$, and free
operations, i.e., $\sfree{a}{c}$ which frees the $c$-colored memory
region allocated at address $a$.
%
%
Lastly, we define the transition system of the monitor over shadow
memories and event history $\stk{\alpha}$ through the judgment
$\rmred{\stk{\alpha}}{T}{\alpha}{T'}$ (\cref{fig:mon}).

\Cref{tr:ms-read,tr:ms-write} consume events
$\com{\rd{a}{(c,s)}}$ and $\com{\wrt{a}{(c,s)}}$, respectively,
provided that the color and the shade are equal to those stored at
location $a$ in shadow memory and that location $a$ is
allocated, i.e., $\com{\absh( a )} = \com{A(c, s)}$.
%
If the colors or the shades do not match, or the memory location is
free, the state machine simply gets stuck, thus detecting a memory
violation.
To consume event $\com{\sallocgen{n}{a}{c}{\phi}}$, \cref{tr:ms-alloc}
allocates \com{n} contiguous, currently \emph{free} locations in
shadow memory, starting at address \com{a}, and assigns fresh color
\com{c} and the shade given by \com{\phi} to them.
%
%
%
%
%
In response to event $\sfree{a}{c}$, the monitor frees the
\com{c}-colored region of memory previously allocated at address
\com{a} through \cref{tr:ms-free}.
First, the rule checks that a matching allocation event is present in
the history, i.e.,
$\stk{\alpha_1} \cdot \sallocgen{n}{a}{c}{\phi} \cdot \stk{\alpha_2}$
for some size $n$ and shading function $\phi$, and that region has not
already been freed, i.e., $\sfree{a}{c} \not\in \stk{\alpha_2}$, and
then sets the tag of the memory locations colored $c$ as free.

%

%
%
We say a trace is memory safe, written $\com{\memsaf{\alsc}}$, if and
only if the state machine does not get stuck while processing the
trace starting from the empty shadow memory $\emptyset$ and empty
history $\epsilon$.
In the definition below, we write $\rmstofat{\alsc}$ for the reflexive
transitive closure of $\rmsto{\ac}$, which accumulates single events
into a trace and records the event history.
%
%
\begin{definition}[Memory Safety]\label{def:ms}
    $\com{\memsaf{\alsc}} \isdef
      \exists \com{\absh}\ldotp \rmredfat{\epsilon}{\emptyset}{\alsc}{\absh}$
\end{definition}
\myfig{
\begin{center}
  \typerule{MS-Read}
{
  T( a ) = A(c,s)
}
{
  \rmred{\stk{\alpha}}{T}{\rd{a}{(c,s)}}{T}
}{ms-read}
\typerule{MS-Write}
{
  T( a ) = A(c,s)
}
{
  \rmred{\stk{\alpha}}{T}{\wrt{a}{(c,s)}}{T}

}{ms-write}
\and
\typerule{MS-Alloc}
{
  \fresh{c}
  &
  \forall j \in \{ 0 ..  n-1 \}\ldotp T(a+j) = F(\_,\_)
  &
  T' = T[a + i \mapsto A(c, \phi(i))  \mid  i \in \{ 0 .. n-1 \} ]
}
{
  \rmred{\stk{\alpha}}{T}{\sallocgen{n}{a}{c}{\phi}}{T'}
}{ms-alloc}
\and
\typerule{MS-Free}
{
  \sfree{a}{c} \not\in \stk{\alpha_2}
  &
  T' = T[i \mapsto F(c,s_i) \mid i \mapsto A(c, s_i) \in T]
}
{
  \rmred{\stk{\alpha_1} \cdot \sallocgen{n}{a}{c}{\phi} \cdot \stk{\alpha_2}}{T}{\sfree{a}{c}}{T'}
}{ms-free}
%
\end{center}
}{mon}{
  Trace-based definition of memory safety.
}


\subsection{Memory Safety of \mswasm.}\label{sec:ms-in-mswa}
%
In order to establish memory safety for \mswasm, we first need to map
the trace model of \mswasm to the abstract trace model of \Cref{sec:ms}.
The main difference between the two is that the abstract model
identifies safe memory accesses using colors and shades, while
\mswasm relies on bounds checks and segment identifiers.
Furthermore, individual $\rdt{h}$ and $\wrtt{h}$ events correspond to
multiple memory accesses in the abstract trace model, as these
operations read and write byte sequences in \mswasm.
We reconcile these differences between the two trace models with the
relation $\traceeq{\at}{}{\trgdelta{}}{\alsc}$ whose most relevant rules are defined in
\cref{fig:tracerel}. 
The relation is parametrized by a partial bijection
$\trgdelta{} : \trg{\mathbb{N}\times \mathbb{N}} \rightharpoonup
\com{\mathbb{N} \times \mc{C} \times \mc{S}}$, which maps pairs
\trg{(a,\tid)}, consisting of an allocated segment memory address \trg{a} and
a segment identifier \trg{\tid}, into corresponding shadow memory addresses
\com{a^{(c, s)}}, decorated with colors and shades.
%
%
Intuitively, we can construct a suitable bijection $\trgdelta{}$ from
the \mswasm allocator, which has information about what is allocated in segment memory.
%
%
%

%
\myfig{\centering
    \typerule{Trace-Read}{
      \trg{h} = \conf{n_b, n_o, \_, \_,\tid}_{\genhandletag}
  		&
  		\trgdelta{}(\trg{n_b,\tid}) = b^{(c,s)}
  		&
  		a = b + \trg{n_o}
      &
  		n = \trg{\sizeof{\trgb{\tau}}} 
  	}{
  		\traceeq{
  			\readt{h}
  		}{}{\trgdelta{}}{
  			\rd{a}{(c,s)}\ \cdots\ \rd{(a+n-1)}{(c,s)}
  		}
  	}{tr-read}
	\and
    \typerule{Trace-Write}{
      \trg{h} = \conf{n_b, n_o, \_, \_,\tid}_{\genhandletag}
	    &
	    \trgdelta{}(\trg{n_b,\tid}) = b^{(c,s)}
	    &
	    a = b + \trg{n_o}
      &
	    n = \trg{\sizeof{\trgb{\tau}}} 
  	}{
  		\traceeq{
  			\writet{h}
  		}{T}{ \trgdelta{}}{
  			\wrt{a}{(c,s)}\ \cdots\  \wrt{(a+n-1)}{(c,s)}
  		}
  	}{tr-write}
  \and
  \typerule{Trace-SAlloc}{
      \trg{h} = \trg{\conf{n_b,0,n_o,\_,\tid}_{\genhandletag}}
      &
        n = \trg{n_o} 
        &
        \forall i \in \{0..n-1\} \ldotp
        \trgdelta{}(\trg{n_b} + i,\trg{\tid}) = (a + i)^{(c, \phi(i))}
    }{
        \traceeq{
            \trg{salloc(h)}
        }{T}{\trgdelta{}}{
            \sallocgen{n}{a}{c}{\phi}
        }
    }{tr-salloc}
    \and
  \typerule{Tr-Sfree}{
    \trg{h} = \trg{\conf{n_b,\_,n,\_,\tid}}
    &
    \trgdelta{}(\trg{n_b},\trg{\tid}) = a^{(c,\_)}
  }{
    \traceeq{\trg{sfree(h,\tid)}}{}{\trgdelta{}}{\sfree{a}{c}}
  }{tr-sfree}
  \and
    \typerule{Trace-Trap}{~}{
    	\traceeq{\trg{\trapt}}{T}{\trgdelta{}}{ \epsilon}
    }{tr-trap}
}{tracerel}{
	Relation between \mswasm and abstract events (excerpts).
}
\Cref{tr:tr-read,tr:tr-write} relate single \mswasm
$\readt{h}_{\genhandletag}$ and
$\writet{h}_{\genhandletag}$ events to a sequence of
$\trg{\sizeof{\trgb{\tau}}}$ contiguous abstract read and write events,
respectively.
%
%
The rules convert the handle base address and the segment identifier
into the corresponding colored base address, i.e.,
$\trgdelta{}(\trg{n_b,\tid}) = b^{(c,s)}$, which is then incremented
with the offset of the handle to obtain the first abstract location
accessed, i.e., $a = b + \trg{n_o}$, similar to \mswasm semantics.
Since these abstract events originate from the same handle, the rule
labels their address with the same color \com{c} and shade \com{s}
obtained from the base address of the handle to reflect their
provenance.
%
%
%
If we computed the colors for these addresses using the bijection
$\trgdelta{}$, then they would automatically match the color stored in
the shadow memory and memory safety would hold trivially.
Instead, these addresses are tagged with the provenance color, and therefore
proving memory safety (i.e., stepping using the rules of \Cref{fig:mon}) requires showing that this
color matches the color found in shadow memory, which in turn requires
reasoning about the integrity of the handle and the bounds checks
performed by \mswasm.
A final subtlety of these rules is that they seem to ignore the
integrity flag of the handle.
This is because in \mswasm, only authentic handles can generate read
and write events---reading and writing memory via corrupted handles
results in a \trg{trap} event.

\Cref{tr:tr-salloc} relates the allocation of a $\trg{n_o}$-byte segment
in \mswasm to a corresponding abstract
allocation of the same size, i.e., event $\sallocgen{n}{a}{c}{\phi}$
where $\trg{n_o} = \com{n}$.
In the rule, premise
$\forall i \in \{0,\ldots,n-1\} \ldotp \trgdelta{}(\trg{a} + i) = (\trg{n_b} +
i, \trg{\tid})^{(c, \phi(i))}$ ensure that (i) all the abstract addresses share
the same color $\com{c}$, and (ii) the bijection $\trgdelta{}$ and the
shading function $\com{\phi}$ agree on the shades used for the
segment.
%
%
%
%
%
%
In general, function $\phi$ can be a constant function, when we reason
about memory safety for native \mswasm programs, e.g., to prove that
\mswasm is memory safe in \Cref{thm:rms} below.
Intuitively, \mswasm does not provide an explicit representation for
structured data, therefore it is sufficient to assign the same shade
to all locations of a segment to prove memory safety.
When we use \mswasm as a compilation target however, segments can
store also structured objects (e.g., a struct) in addition to flat
objects (e.g., an array).
In this scenario, we instantiate $\phi$ according to the source type
of the object, which let us show that compiled C/C++ programs achieve
intra-object memory safety later (\Cref{thm:comp-rs-strong}).

To relate free events, \cref{tr:tr-sfree} requires the bijection
\trgdelta{} to match the base \trg{n_b} and identifier \trg{\tid} of
the segment pointed to by the handle to the colored address $a^c$
freed by the monitor, i.e., $\trgdelta{}(\trg{n_b},\trg{\tid}) =
a^{(c,\_)}$.
Because identifiers and colors are never reused, freed segments and
regions can be reused for other allocations, while
keeping dangling handles and colored pointers related by the
bijection.
For example, if a segment is later allocated at address \trg{n_b}, it
will be associated with a \emph{unique} identifier
$\trg{\tid'} \neq \trg{\tid}$, which can be related to some shadow
address $a'$ and \emph{fresh} color $c' \neq c$ through an
\emph{extended} bijection
$\trgdelta' \supseteq \trgdelta{}$.\footnote{In technical terms the
  bijection grows \emph{monotonically}, which provides a suitable
  inductive principle for our formal results.}

%
Lastly, \cref{tr:tr-trap} relates event \trapt in \mswasm to the empty
trace $\epsilon$, since $\trapt$ simply stops the program and thus
cannot cause a memory safety violation.

  We can now state memory safety for \mswasm traces in terms of memory
  safety of a $\trgdelta{}$-related abstract trace for the state
  machine defined above.

\begin{definition}[Memory Safety for \mswasm Traces]\label{def:ms-mswasm}
	\(
	\trgms{\alst} \isdef
		\exists \alsc, {\trgdelta{}} \ldotp
		\traceeq{\trg{\alst}}{}{ {\trgdelta{}}{}}{\alsc} \text{ and } \memsaf{\alsc}
	\)
\end{definition}
We define memory safety for \mswasm modules if the trace generated
during execution is memory safe.
In the following, we write $\trg{M \semt \alst}$ for the trace
generated by module $\trg{M}$ with the
semantics of \Cref{sec:sem}.

%
\begin{definition}[Memory Safety for \mswasm Modules]\label{def:mod-ms}
  \(
    \vdash\trgms{\trg{M}} \isdef
      \trg{M \semt \alst}
      \text{ and }
      \trgms{\alst}
  \)
\end{definition}


A module \trg{M} achieves robust memory safety if, given any valid attacker \trg{\ctx} (denoted as $\trg{M}\vdash \trg{\ctx} : \trg{attacker}$, in the sense of \Cref{sec:atk-model}), linking \trg{M} with \trg{C} produces a memory
safe module.
In the following, we write $\trg{M\composition \ctx}$ for the module
obtained by linking \trg{M} with \trg{C}, i.e., the module obtained
by instantiating the functions imported by \trg{M} with those
of \trg{C}.
%

\begin{definition}[Robust Memory Safety for \mswasm Modules]\label{def:mod-ms-rob}
  \(
    \vdash\rms{\trg{M}} \isdef
      \forall \trg{\ctx} \text{ s.t. } \trg{M}\vdash \trg{\ctx} : \trg{attacker}\ldotp
      \vdash\trgms{\compose{\trg{M}}{\trg{\ctx}}}
  \)
\end{definition}

The main result for \mswasm is that any well-typed module ($\vdash\trg{M}:\trg{wt}$) is memory-safe, robustly.
\begin{theorem}[Robust Memory Safety for \mswasm]\label{thm:rms}
    \(
    \text{ If } \vdash\trg{M}:\trg{wt} \text{ then } \vdash\rms{\trg{M}}
    \)
\end{theorem}
\begin{proofsketch}
  Intuitively, the type system of \mswasm ensures that well-typed
  modules can access segment memory only through \trg{\handle} values
  and safe instructions.
  Programs that accesses memory via \emph{invalid} handles
  \trg{\trapt} and so trivially respect memory safety
  (\Cref{tr:tr-trap}).
  When accessing segments via \emph{valid} handles, \mswasm performs
  memory safety checks using their metadata, so the rest of the proof
  requires showing an invariant about handle integrity.
  Intuitively, this invariant guarantees that valid handles (whether
  proper values, or stored in segment memory), correspond to allocated
  segments in memory.
  Then, using this invariant, we can show that programs that access
  segment memory without trapping, pass the memory safety checks, and
  thus are memory safe.
\end{proofsketch}


In the next section, we leverage the memory-safety abstractions of
\mswasm to develop a formal \clang compiler that provably enforces
memory safety.




%

\section{Memory Safety Through Compilation}\label{sec:compiler}
This section shows how a \clang compiler targeting \mswasm can enforce memory safety.
Thus, we formalize a simplified version of \clang
(\Cref{sec:src}), as a memory-unsafe source language for our compiler,
and the compiler itself (\Cref{sec:comp}).
We then prove that the compiler enforces memory safety
(\Cref{sec:comp-prop}), i.e., memory-safe programs compiled to \mswasm
execute unchanged (\Cref{thm:comp-rs-strong}), while memory-unsafe
programs abort at the first memory violation (\Cref{thm:trap}).
%



\subsection{The Source Language \clang}\label{sec:src}

\Cref{fig:c-syntax} presents the syntax of our source language, a
subset of \clang, inspired by previous work~\cite{checkedc}.
Source programs \src{M} specify the type of imported functions
\src{\stk{I}}, struct definitions \src{\stk{D}}, function definitions
\src{\stk{F}}, and the heap size \src{n_{hs}}.
%
%
%
Struct definitions map struct names \src{s} to a list of field names
\src{f} and their types \src{\tau}.
Types are mutually defined by expression types \src{\tau}, i.e.,
integers (\src{int}) and pointers (\src{\ptrc{w}}), and word types
\src{w}, which, in addition to \src{\tau}, include also multi-word
values, i.e., structs (\src{\struct{s}}) and arrays
(\src{\arr{n}{\tau}}).
Syntax $\src{\tau\ g (x: \tau)\ \{\ \srcvar\ \stk{(y:\tau)}, e \}}$
defines function $\src{g}$, its argument and return type, and declares
local variables $\src{\stk{(y:\tau)}}$ in scope of the body $\src{e}$.
Expressions are standard and include reading and writing memory via
pointers (i.e., $\src{\drf{e}}$ and $\src{\asgn{e}{e}}$) and accessing
struct fields (i.e., $\src{\lkup{e}{f}}$).
Expression $\src{\pmalloct{\tau}{e}}$ allocates an array containing
$\src{e}$ elements of type $\src{\tau}$, while $\src{\pmalloc{w}}$
allocates a buffer to store a single element of type
$\src{w} \neq \src{array\ \tau}$.
%
%
%
%
Values include integers \src{n} and annotated pointers, i.e.,
$\src{n^{(n_1,n_2,w,\tid)}}$, where $\src{n}$ is the address pointed to by
the pointer, and $\src{(n_1,n_2,w,\tid)}$ indicates that the pointer refers
to a buffer allocated at address $\src{n_1}$, containing $\src{n_2}$
elements of type $\src{w}$, and identified by \src{\tid}.
These annotations are inspired by previous work on pointer
provenance~\cite{pointer-provenance} and are only needed to reason
about memory safety of source programs, i.e., the source semantics
does \emph{not} enforce memory safety and ignores them.

%
\myfig{\centering
\begin{gather*}
  \begin{aligned}
    \printnames{\text{Programs } }
    \src{M} \bnfdef
      &\
      \src{\stk{I}, \stk{D}, \stk{F}, n_{hs}}
    &
    \printnames{\text{Imports } }
    \src{I} \bnfdef
      &\
      \src{\tau\ g(x:\tau)}
    &
    \printnames{\text{Structs }}
    \src{D}\bnfdef
    	&\
    	\src{s\mapsto \stk{(f:\tau)}}
    \end{aligned}
    \\
    \begin{aligned}
    \printnames{\text{Values }}
    \src{v} \bnfdef
      &\
      \src{n} \mid \src{n^{(n,n,w,\tid)}}
    &
      \printnames{\text{Functions } }
      \src{F} \bnfdef
        &\
        \src{\tau\ g (x: \tau)\ \{\ \srcvar\ \stk{(y:\tau)}, e \}}
      \\
      \printnames{\text{Word Types } }
      \src{w} \bnfdef
        &\
        \src{\tau} \mid \src{\struct{s}} \mid \src{\arr{n}{\tau}}
      &
      \printnames{\text{Expr. Types } }
      \src{\tau} \bnfdef
        &\
        \src{int} \mid \src{\ptrc{w}}
    \\
    \end{aligned}
    \\
    \begin{aligned}
    \printnames{\text{Expr. } }
    \src{e} \bnfdef
    	&\
		\src{v} \mid\src{x} \mid \src{\seq{e}{e}} \mid \src{e\op e} \mid
                \src{\varasgn{x}{e}} \mid \src{\letcall{x}{g}{e}{e}} \mid \src{\drf{e}}
    \\
    \mid&\
    \src{\asgn{e}{e}} \mid \src{\ifte{e}{e}{e}} \mid \src{\lkup{e}{f}} \mid \src{\pmalloct{\tau}{e}}
    \mid
      \src{\pmalloc{w}} \mid \src{\freesrc{e}}
    \end{aligned}
    \\
    \begin{aligned}
       \printnames{\text{Stores } }
	\src{\Sigma} \bnfdef&\
			\srcconf{H, A}
       &
       \printnames{\text{Heaps } }
  \src{H} \bnfdef&\
    \src{\nil} \mid \src{ v \stackcat H }
       &
       \printnames{\text{Local Env. } }
    \src{\theta} \bnfdef&\
      \src{\stk{(x \mapsto v)}}
       &
      \printnames{\text{Allocators }}
        \src{A}
        &\
	\end{aligned}
	\\
	\begin{aligned}
	       \printnames{\text{Events } }\as \bnfdef
			&\
			\src{\epsilon} \mid \src{alloc(v)} \mid \src{\frees{v}} \mid \src{\psrcrd{\tau}{v}} \mid \src{\psrcwrt{\tau}{v}}
	\end{aligned}
\end{gather*}
}{c-syntax}
{ \clang syntax and runtime structures (excerpts). }

\myparagraph{Typing.}
The type system for the source language is mostly standard and defined
by judgment $\src{\stk{F},\Gamma\vdash e:\tau}$, which indicates that
expression \src{e} has type \src{\tau} under functions \src{\stk{F}}
and typing context \src{\Gamma} (which binds variables to types).
The type system allows typing integers as pointers and restricts
function type signatures to expression types for simplicity.
%

\myparagraph{Semantics.}
We define a small-step contextual semantics for \clang with the following judgment,
$\srcstepprimconfig{\srcconfig{\Sigma}{\theta}{e}}{\alpha}{\srcconfig{\Sigma'}{\theta'}{e'}}$, in which local
configuration \src{\srcconfig{\Sigma}{\theta}{e}} steps and produces event
\as, under program definition \src{M}.
Local configurations contain the store \src{\Sigma}, the local
variable environment \src{\theta} mapping named variables to values,
and an expression \src{e} to be evaluated.
The store \src{\Sigma} contains the heap \src{H}, a list of values, and the allocator state \src{A}.
The heap abstracts away low-level details about the memory layout and the byte
representation of values (e.g., we store structs and arrays simply as a
flattened sequence of single-word values).
%
%
%
%
Similar to \mswasm, events \as record memory relevant operations,
including silent events $\src{\epsilon}$, allocating and releasing
memory, i.e., $\src{alloc(v)}$ and $\src{\frees{v}}$, and reading and
writing values of type \src{\tau} with a pointer \src{v}, i.e.,
\src{\psrcrd{\tau}{v}} and \src{\psrcwrt{\tau}{v}}.
%



\myfig{\centering
	\typerule{Ptr-Arith}{
	}{
		\srcstepprimconfig{ \srcconfig{\Sigma}{\theta}{ \aptr{a}{b}{\ell}{w} \op n}}{}{ \srcconfig{\Sigma}{\theta}{\aptr{(a+n)}{b}{\ell}{w}}}
	}{src-ptr-arith}
	\and
	\typerule{Write-Ptr}{
                \src{\Sigma} = \srcconf{H, A} &
		\src{H'} = \src{H[a \mapsto v]} &
                \src{\Sigma'} = \srcconf{H', A} &
                \src{\vdash v : \tau}
	}{
		\srcstepprimconfig{ \srcconfig{\Sigma}{\theta}{\asgn{\aptr{a}{b}{\ell}{w}}{v}}}{{\psrcwrt{\tau}{\aptr{a}{b}{\ell}{w}}}}{\srcconfig{\Sigma'}{ \theta}{\src{0}}}
	}{src-assign}
	\and
	\typerule{Write-Int}{
                \src{\Sigma} = \srcconf{H, A} &
		\src{H'} = \src{H[a \mapsto v]} &
                \src{\Sigma'} = \srcconf{H', A} &
                \src{\vdash v : \tau}
	}{
		\srcstepprimconfig{ \srcconfig{\Sigma}{\theta}{\asgn{a}{v}}}{{\psrcwrt{\tau}{a}}}{\srcconfig{\Sigma'}{ \theta}{ \src{0}}}
	}{src-assign-forge}
	\and
	\typerule{Malloc-Single}{
            \srcastep
              {\Sigma = \srcconf{H, A}}
              {\src{alloc(\aptr{a}{a}{1}{w})}}
              {\srcconf{H', A'} = \Sigma'}
              &
              \src{v} = \src{\aptr{a}{a}{1}{w}}
             }{
               \begin{multlined}
  \srcstepprimconfig{ \srcconfig{\Sigma}{\theta} {\pmalloc{w}}}
                    {{\src{alloc(v)}}}
                    {\srcconfig{\Sigma'}{ \theta}{v}}
               \end{multlined}
  }{src-malloc-single}
  \and
	\typerule{Malloc-Array}{
            \srcastep
              {\Sigma = \srcconf{H, A}}
              {\src{alloc(\aptr{a}{a}{n}{\tau})}}
              {\srcconf{H', A'} = \Sigma'}
              &
              \src{v} = \src{\aptr{a}{a}{n}{\tau}}
             }{
               \begin{multlined}
  \srcstepprimconfig{ \srcconfig{\Sigma}{\theta} {\pmalloct{\tau}{n}}}
                    {{\src{alloc(v)}}}
                    {\srcconfig{\Sigma'}{ \theta}{v}}
               \end{multlined}
  }{src-malloc-array}
}{c-sem}{
    Semantics of \clang (excerpts).
}

\Cref{fig:c-sem} presents some of the semantics rules of the source language.
%
%
\Cref{tr:src-ptr-arith} performs pointer arithmetic by incrementing
the address of the pointer, without changing the metadata.
\Cref{tr:src-assign,tr:src-assign-forge} write a value \src{v} in the
heap through a pointer and a raw integer address, respectively.
As explained above, these rules do not check that the write operation
is safe, but only record the pointer and the type \src{\tau} of the
value that gets stored in the generated event, i.e.,
$\psrcwrt{\tau}{\aptr{a}{b}{\ell}{w}}$ and $\psrcwrt{\tau}{a}$.
\Cref{tr:src-malloc-single,tr:src-malloc-array} allocate a buffer for
a single object of type $\src{w}$ and a $\src{n}$-elements array,
respectively, and return a pointer value annotated with appropriate
metadata.
Similar to the \mswasm semantics, the source language invokes the
allocator to serve allocation and free requests ($\srcastep
              {\srcconf{H, A}}
              {\src{\as}}
              {\srcconf{H', A'}}$).
In contrast to the safe allocator of \mswasm however, the source
allocator does not trap upon an invalid free request, i.e., a free of
an unallocated memory region, but silently drops the
request.\footnote{Invalid free requests cause \emph{undefined
    behavior} in C and usually result in the corruption of memory
  objects or the allocator state. Since we represent the allocator
  state explicitly and separately from the program memory, free
  requests cannot cause such specific behaviors in our model. }

%
%


The source language uses a separate semantic judgment for function
calls and returns (omitted), and a top-level judgment \src{M \srcsemt \alss},
which collects the trace generated by program
\src{M}.
We define memory safety for source traces using the general abstract monitor
from \Cref{sec:ms}:
\begin{definition}[Memory-Safety for \clang]
	\(
	\srcms{\alss} \isdef
		\exists \alsc, {\srcdelta{}} \ldotp
		\traceeq{\src{\alss}}{}{ {\srcdelta{}}{}}{\alsc} \text{ and } \memsaf{\alsc}
	\)
\end{definition}
This definition is analogous to \Cref{def:ms-mswasm} for \mswasm: it relies on a
bijection $\srcdelta{}$ to map source addresses into corresponding colored abstract
addresses, and a relation $\traceeq{\src{\alss}}{}{ {\srcdelta{}}{}}{\alsc}$
to connect source and abstract traces through the bijection.
%
The trace relation is defined similarly to the relation given in
\Cref{fig:tracerel} for \mswasm and additionally constructs
appropriate shading functions for \src{alloc(v)} events according to
the type of the allocated object (e.g., an array or a struct).
%
Accesses via raw addresses $\src{n}$ are excluded from
the relation, i.e.,
$\traceneq{\psrcwrt{\tau}{a}}{}{\srcdelta{}}{\alsc}$ and
$\traceneq{\psrcrd{\tau}{a}}{}{\srcdelta{}}{\alsc}$ for any abstract
trace $\alsc$.
Omitting them from the relation captures the fact that
memory accesses with forged pointers violate memory safety, as the provenance
of these pointers is undefined.
%
%

\subsection{The Compiler}\label{sec:comp}
We define the compiler \comp{\cdot} from
\clang to \mswasm inductively on the type derivation of \clang modules, functions and expressions (\Cref{fig:comp}).
To prevent untrusted code from violating memory safety, our compiler
translates pointers to handles and only uses \mswasm segment memory.
Thus, we translate source types \src{\tau} into \mswasm types
\trgb{\tau} as $\compct{\src{int}}{\trg{i32}}$ and
$\compct{\src{\ptrc{w}}}{\trg{handle}}$.
%
%
The compiler relies on source types to emit \mswasm instructions with
appropriate byte sizes (calculated with function $\sz{\cdot} : \src{\tau}\to\trg{n}$) and offsets for expressions that involve
pointer arithmetic, struct accesses and memory allocations.
%
%
%
%
\myfig{\centering
    \typerule{C-Ptr-BinOp}{
           \compc{\src{f}}{ \strule{P,\Gamma}{e_1}{\arr{}{\tau}}}{\trg{i^*_1}}
           &
           \compc{\src{f}}{ \strule{P,\Gamma}{e_2}{int}}{\trg{i^*_2}}
           &
           \trg{n} = \sz{\src{\tau}}
      }{
          \compc{\src{f}}{\strule{P,\Gamma}{e_1 \op e_2}{\arr{}{\tau}} }{\trg{i^*_1; i^*_2; i32.const\ n; i32.\times; \add}}
      }{c-binop-pt}
           \and
  \typerule{C-BinOp}{
           \compc{\src{f}}{ \strule{P,\Gamma}{e_1}{w}}{\trg{i^*_1}}
           &
           \compc{\src{f}}{ \strule{P,\Gamma}{e_2}{w}}{\trg{i^*_2}}
           &
           \compct{\src{\tau} }{\trgb{\tau}}
      }{
          \compc{\src{f}}{\strule{P,\Gamma}{e_1 \op e_2}{\tau} }{\trg{i^*_1; i^*_2;\bopt}}
      }{c-binop}
           \and
    \typerule{C-Malloc-Array}{
    	\comp{\src{\Gamma}\vdashm\src{e}:\src{int} }={\trg{\stk{i_1}}}
    	&
        &
    	&
    	\trg{n} = \sz{\src{\tau}}
    }{
    		\comp{\src{\Gamma}\vdashm\src{\pmalloct{\tau}{e}}:\src{\ptrc{(array\ \tau)}} } =
    		\trg{ \stk{i_1}; \ittw.const~n; \ittw.\otimes; \newst }
    }{c-malloc-array}
    \and
          \typerule{C-Malloc-Single}{
              \trg{n} = \sz{\src{w}}
          }{
              \compc{
                \src{f}
              }{
                \strule{P,\Gamma}{\pmalloc{w}}{\ptrc{w}}
              }{
                \trg{ i32.const\ n; \newst}
              }
          }{c-malloc-single}
           \and
    \typerule{C-Deref}{
    	\compct{\src{\Gamma}\vdashm\src{e}: \src{\ptrc{\tau}} }{\trg{\stk{i}}}
    	&
    	\compct{\src{\tau} }{\trgb{\tau}}
    }{
    	\compct{\src{\Gamma}\vdashm\src{\drf{e}}: \src{\tau} }{\trg{\stk{i}\stackapp [\hload]}}
    }{c-deref}
	\and
	\typerule{C-Struct-field}{
		\compct{\src{\Gamma}\vdashm\src{e}: \src{\ptrc{(\struct{s})}} }{\trg{\stk{i}}}
		&
		\trg{(o_1, o_2)} = \fun{offset}{\src{s}, \src{f}}
	}{
			\comp{\src{\Gamma}\vdashm\src{\lkup{e}{f}}: \src{\ptrc{\tau}} } =
			\trg{\stk{i} \stackapp \trg{[\ittw.const\ o_1 , \ittw.const\ o_2 , \hslice]} }
	}{c-struct-field}
}{comp}{
  Compiler from \clang to \mswasm (excerpts).
}
%
%
For example, a binary operation (\src{\op}) whose first operand is an array (\src{\arr{}{\tau}}) needs to be compiled in a \trg{\add}, as in \Cref{tr:c-binop-pt}.
On the other hand, a binary operation on naturals needs to be compiled in the related \mswasm binary operation, as in \Cref{tr:c-binop}.
Another example of the way source types guide the compilation is for the compilation of expression \src{\pmalloct{\tau}{e}}.
Here, if the resulting type is a pointer to an array (\src{\ptrc{(array\ \tau)}}), the
compiler must first emit instructions to compute the size of a segment
large enough for an array containing \src{e} elements of type
\src{\tau}, and then instruction $\newst$ to invoke the
allocator and generate the corresponding handle.
Therefore, \cref{tr:c-malloc-array} recursively compiles the array
length \src{e}, i.e.,
$\comp{\src{\Gamma}\vdashm\src{e}:\src{int} }={\trg{\stk{i}}}$, which
then gets multiplied by \abs{\src{\tau}}, i.e., the size in bytes of a
value of type \src{\tau}, via instruction \trg{\ittw.\otimes}, and
finally passed to $\newst$.
%
%
On the other hand, if the return type is a pointer to any other type (\src{\ptrc{w}}), the compiler needs to calculate its size (\trg{n}) and allocate enough memory (\Cref{tr:c-malloc-single})
Since expression \src{\drf{e}} reads a pointer to a value of type
\src{\tau}, \cref{tr:c-deref} emits instruction to first evaluate the
corresponding handle, i.e.,
$\compct{\src{\Gamma}\vdashm\src{e}: \src{\ptrc{\tau}}
}{\trg{\stk{i}}}$, followed by instruction \hload, whose compiled type
$\compct{\src{\tau} }{\trgb{\tau}}$ ensures that the generated code
reads the right number of bytes and interprets them at the
corresponding target type.
%
%
Lastly, \cref{tr:c-struct-field} translates a struct field access
\src{\lkup{e}{f}} by slicing the handle obtained from pointer \src{e},
thus enforcing intra-object safety in the generated code.
To this end, the rule emits instructions \trg{[\ittw.const\ o_1 ,
  \ittw.const\ o_2 , \hslice]}, where offsets \trg{o1} and \trg{o2}
are obtained from function \fun{offset}{\src{s}, \src{f}}, which
statically computes the offsets necessary to select field \src{f} in
the byte representation of struct \src{s}.
%


\subsection{Properties of the Compiler}\label{sec:comp-prop}
We establish two properties for our compiler.
The first (\Cref{thm:comp-rs-strong}) shows that the compiler is functionally
correct and preserves memory safety for memory-safe source programs.
%
%
The second (\Cref{thm:trap}) shows that \emph{memory-unsafe} programs
compiled to \mswasm abort at the first memory violation.
Together, these results show that our compiler enforces memory-safety (\Cref{thm:cor-enf}).

\myparagraph{Cross-Language Equivalence Relation.}
Since our notion of memory safety is defined over traces, and the source
and target languages have different trace models, the formal results
of the compiler rely on a \emph{cross-language} equivalence relation
to show functional correctness and memory-safety preservation~\cite{Leroy09b}.
%
%
\Cref{fig:cross-rel} (top) defines this relation for pointer values up to a partial bijection
$\crossdelta{} : \src{\mathbb{N}\times \mathbb{N}} \rightharpoonup
\trg{\mathbb{N}\times \mathbb{N}}$, which maps addresses and
identifiers from source to target.
\Cref{tr:val-rel-ptr} relates an annotated pointer
$\src{\aptr{a}{b}{\ell}{w}}$ to a \emph{valid} handle
$\trg{\conf{b, o, \ell, true, \tid}}$ as long as their base address
and identifier are matched by the bijection, i.e.,
$\crossdelta{}(\src{b,\tid}) = \trg{b,\tid}$, and the length and
offset fields match, taking into account the byte-size representation
of $\src{w}$, i.e., $\src{\ell}\times \abs{\src{w}} = \trg{\ell}$ and
$\src{(a-b)} \times \abs{\src{w}} = \trg{o}$.
In contrast, \Cref{tr:val-rel-int} relates integer pointers to arbitrary
\emph{invalid} handles.
The relation between source and target events
$\relatesrctrgtr{\src{\alpha}}{\crossdelta{}}{\trg{\alpha}}$, relates the same single events (\Cref{fig:cross-rel}, bottom).
When relating reads, writes, allocates, and frees, we insist that source pointers and target handles are related (according to the cross-language value relation) and the handles are valid (these rules have the validity bit set to \trg{true}).
Additionally, for reading and writing, they values being read or written must be of related types, i.e., $\compct{\src{\tau}}{\trgb{\tau}}$.

\myfig{\centering
\centering
    \typerule{Val-Rel-Ptr}{
      \crossdelta{}(\src{b,\tid}) = \trg{b,\tid}
      &
      \src{\ell}\times \abs{\src{w}} = \trg{\ell}
      &
      \src{(a-b)} \times \abs{\src{w}} = \trg{o}
    }{
        \compd{}{\src{\aptr{a}{b}{\ell}{w}}}{{\conf{b, o, \ell, true,\tid}}}
    }{val-rel-ptr}
    \and
    \typerule{Val-Rel-Int}{
    }{
        \compd{}{\src{n}}{{\conf{b, o, \ell, false,\tid}}}
    }{val-rel-int}
%
\smallskip
\hrule

\begin{center}
    \typerule{Tr-Rel-
    Read-Ptr}{
        \compct{\src{\tau}}{\trg{\tau}}
        &
        \compd{}{\src{\aptr{a}{b}{\ell}{w}}}{\trg{\conf{b, o, \ell,true,\tid}}}
    }{
        \relatesrctrgtr{\psrcrd{\tau}{\aptr{a}{b}{\ell}{w}}}{\crossdelta{}}{\readt{\conf{b, o, \ell, true, \tid}}}
    }{trrel-read}
    \and
    \typerule{Tr-Rel-Write-Ptr}{
        \compct{\src{\tau}}{\trg{\tau}}
        &
        \compd{}{\src{\aptr{a}{b}{\ell}{w}}}{\trg{\conf{b, o, \ell, true, \tid}}}
    }{
        \relatesrctrgtr{\psrcwrt{\tau}{\aptr{a}{b}{\ell}{w}}}{\crossdelta{}}{\writet{\conf{b, o, \ell, true}}}
    }{trrel-write}
 \typerule{Tr-Rel-Allocate}{
           \compd{}{\src{(\aptr{a}{b}{\ell}{w})}}{\trg{\conf{b,o,\ell,true,\tid}}}
         }
         {\relatesrctrgtr{\src{salloc(\aptr{a}{b}{\ell}{w})}}{\crossdelta{}}{\trg{salloc(\conf{b,o,\ell,true,\tid})}} }
         {trrel-allocate} \and
 \typerule{Tr-Rel-Free}{
           \compd{}{\src{(\aptr{a}{b}{\ell}{w})}}{\trg{\conf{b,o,\ell,true,\tid}}}
         }
         {\relatesrctrgtr{\src{sfree(\aptr{a}{b}{\ell}{w})}}{\crossdelta{}}{\trg{free(\conf{b,o,\ell,true,\tid})}} }
         {trrel-free}

\end{center}
}{cross-rel}{
	Cross-language equivalence relation: values (top) and events (bottom).
}


%
%

%
For memory-safe, well-typed source programs (\struleprog{M}),
\Cref{thm:comp-rs-strong} states that the compiler produces
equivalent memory-safe target programs; i.e., the compiled program
emits a memory-safe trace that is related to the source trace.

\begin{theorem}[Memory-Safety Preservation]\label{thm:comp-rs-strong} 
\begin{align*}
	\text{If }
		&
		\struleprog{M}
		\text{ and }
		\src{M \srcsemt \alss} 
		\text{ and } \srcms{\alss}
	\text{ then }
    \exists \crossdelta{}, \alst \ldotp
    \comp{\src{M}} \trg{\semt \alst}
  	\text{ and }
		\relatesrctrgtr{\alss}{\crossdelta{}}{\alst}
		\text{ and }
		\trgms{\alst}
\end{align*}
\end{theorem}

%


\noindent
In contrast, \Cref{thm:trap} states that memory-unsafe programs compiled to
\mswasm abort at the first memory violation.

%
\begin{theorem}[Memory Violations Trap]\label{thm:trap}
\begin{align*}
	\text{If }
		&
		\struleprog{M}
		\text{ and }
    \src{M \srcsemt \alss \stackapp [\as] \stackapp \alssp}
		\text{ and }
		\srcms{\alss}
	\text{ and }
			\neg \srcms{\src{\alphaseq \stackapp [\as]}}
	\\
	\text{then }
		&
    \exists \crossdelta{}, \alst \ldotp
    \relatesrctrgtr{\alss}{\crossdelta{}}{\alst}
    \text{ and }
    \comp{\src{M}} \trg{\semt \alst \stackapp [\trapt]}
\end{align*}
\end{theorem}

\noindent
Together these theorems characterize the scope of our compiler-based
memory-safety \emph{enforcement}:


\begin{corollary}[Memory-Safety Enforcement]\label{thm:cor-enf}
  \(
    \text{ If } \comp{\src{M}} \trg{\semt \alst} \text{ then } \trgms{\alst}
  \)
\end{corollary}





\Cref{fig:proof} shows the essence of the proof technique for \Cref{thm:comp-rs-strong}, in the diagram, full arrows represent hypotheses and dashed arrows represent conclusions.

\myfig{
\begin{center}
\begin{tikzpicture}[line cap=round,line join=round,x={(-0.5cm,-0.5cm)},y={(1cm,0cm)},z={(0cm,1cm)}]
	\tikzstyle{state} = [fill=white,circle,minimum size=1pt,inner sep=1pt]
	\tikzstyle{nodeup} = [midway,sloped,above,font=\footnotesize]
	\tikzstyle{nodedn} = [midway,sloped,below,font=\footnotesize]

	\begin{pgfonlayer}{main}
		\def\b{2.5}
		\def\L{4}
		\def\a{60} 
		\foreach\i in {0,1}
		{
			\begin{scope}[canvas is yz plane at x=-\L*\i]
			\coordinate (A\i) at (-1,0);
			\coordinate (B\i) at (-\b,.5);
			\coordinate (C\i) at (190-\a:\b);
			\end{scope}
		}
	\end{pgfonlayer}

	\begin{pgfonlayer}{foreground}
		\node[state]at(A0)(a0w){\trg{A}};
		\node[state]at(A1)(a1w){\trg{A'}};
		\node[state]at(B0)(b0w){\src{A}};
		\node[state]at(B1)(b1w){\src{A'}};

		\draw[->,color = \stlccol, thick] (b0w.north east) to node[nodeup,pos = .5](as){\as} (b1w.south west);
		\draw[->,dashed, color = \ulccol, thick] (a0w.north east) to node[nodeup,pos = .5](as){\at} (a1w.south west);
		\draw[-] (b0w) to node[nodedn](td){} (a0w);
		\draw[-,dashed] (b1w) to node[nodeup](tdp){} (a1w);


		\draw[fill=white,draw=none,opacity=0.4] (A0) -- (A1) -- (B1) -- (B0) -- (A0);
	\end{pgfonlayer}

	\begin{pgfonlayer}{background}
		\node[below=.1 of A0, yshift=-2em,state](s0t){\trgb{\Omega}};
		\node[below=.1 of A1, yshift=-2em,state](s1t){\trgb{\Omega'}};
		\node[below=.1 of B0, yshift=-2em,state](s0s){\src{\Omega}};
		\node[below=.1 of B1, yshift=-2em,state](s1s){\src{\Omega'}};


		\draw[->,color = \stlccol, thick] (s0s.north east) to node[nodeup,pos = .5](as){\as} (s1s.south west);
		\draw[->,dashed, color = \ulccol, thick] (s0t.north east) to node[nodeup,pos = .5](at){\at} (s1t.south west);
		\node[below =.01 of s1t.west,xshift = -.2em,yshift = .3em](){\trg{*}};

		\draw[-] (s0s) to node[nodedn](td){$\hequiv{}{\crossdelta{}}{}$} (s0t);
		\draw[-,dashed] (s1s) to node[nodeup,pos=.3](tdp){$\hequiv{}{\crossdelta{'}}{}$} (s1t);

		\draw[-,dashed] (as) to node[nodeup](tr){$\relatesrctrgtr{}{\crossdelta{'}}{}$} (at);

	\end{pgfonlayer}

	\begin{pgfonlayer}{veryback}
		\draw[fill=black,draw=none,opacity=0.1] (A0|-s0t) -- (A1|-s1t) -- (B1|-s1s) -- (B0|-s0s) -- (A0|-s0t);
	\end{pgfonlayer}
	\begin{pgfonlayer}{main}
		\node[right =7 of a0w ](a0wl){\trg{A}};
		\node[right =7 of a1w ](a1wl){\trg{A'}};
		\node[right =7 of b0w ](b0wl){\src{A}};
		\node[right =7 of b1w ](b1wl){\src{A'}};
		\node[right =7 of C0 ](c0wl){\com{T}};
		\node[right =7 of C1 ](c1wl){\com{T'}};

		\draw[fill=\stlccol!50!white,draw=none,opacity=0.2] (B0-|b0wl) -- (C0-|c0wl) -- (C1-|c1wl) -- (B1-|b1wl) -- (B0-|b0wl);
		\draw[fill=\ulccol!50!white,draw=none,opacity=0.2] (A0-|a0wl) -- (C0-|c0wl) -- (C1-|c1wl) -- (A1-|a1wl) -- (A0-|a0wl);

		\draw[->, thick] (c0wl) to node[nodeup,pos = .5](ac){\ac} (c1wl);
		\draw[->,color = \stlccol, thick] (b0wl) to node[nodeup,pos = .5](as){\as} (b1wl);
		\draw[-] (b0wl) to node[nodeup,midway](sd){$=_\srcdelta{}$} (c0wl);
		\draw[-] (b1wl) to node[nodedn,midway](sdp){$=_\srcdelta{'}$} (c1wl);

		\draw[->, color = \ulccol, thick] (a0wl) to node[nodeup,pos = .5](at){\at} (a1wl);
		\draw[-,dashed] (c0wl) to node[nodedn,pos=.7](td){$=_\trgdelta{}$} (a0wl);
		\draw[-,dashed] (c1wl) to node[midway,sloped,above](tdp){$=_\trgdelta{'}$} (a1wl);

		\draw[-] (b0wl) to node[nodedn,pos=.5](td){$=_{\delta{}}$} (a0wl);
		\draw[-] (b1wl) to node[nodeup,pos=.5](td){$=_{\delta{'}}$} (a1wl);

		\draw[-] (as) to node[nodedn,pos=.5](td){$=_{\delta{'}}$} (at);
		\draw[-] (as) to node[nodedn,pos=.4](td){$=_\srcdelta{'}$} (ac);
		\draw[-,dashed] (at) to node[nodeup,pos=.2](td){$=_\trgdelta{'}$} (ac);
	\end{pgfonlayer}
\end{tikzpicture}
\end{center}
}{proof}{
	Proof diagram for \Cref{thm:comp-rs-strong}: functional correctness (left) and memory-safety preservation (right).
}
%
In the theorem statement, judgements of the form \com{M \sem \alsc} unfold to the reflexive-transitive closure of a single semantics step (i.e., the rules presented in \Cref{fig:mswasm-intra-rules} for \mswasm and in \Cref{fig:c-sem} for \clang).
The proof then proceeds unsurprisingly by induction over the reflexive-transitive reductions that generate the source trace, the figure shows the single-step case. 
We use metavariable \com{\Omega} to indicate program states, which are the tuples presented in the semantics rules of each language. 

%
We first describe the most interesting case of the functional correctness part of \Cref{thm:comp-rs-strong}, i.e., the left of \Cref{fig:proof}.
There, we need to show how one single source step (\src{\Omega\xtoudbl{\as}{}{}\Omega'}) that triggers a change in the source allocator (\srcastep{A}{\as}{A'})%
\footnote{For ease of reading, we massage the allocator reduction judgement $\srcastep
              {\srcconf{H, A}}
              {\src{\as}}
              {\srcconf{H', A'}}$ to only contain the allocator.}, causes a series of `related' target steps (\trg{\trgb{\Omega}'\xtoudbl{\at}{*}{}\trgb{\Omega'}}) that change the target allocator accordingly (\astep{A}{\at}{A'}). 
Essentially, target steps are related when they generate actions that are related (as per \Cref{fig:cross-rel}), and they take related states ($\compd{}{\src{\Omega}}{\trgb{\Omega}}$) into still-related states ($\compd{}{\src{\Omega'}}{\trgb{\Omega'}}$).
We do not present the formalisation of the state relation, intuitively it just lifts the value relation of \Cref{fig:cross-rel} to all elements of a program state.
Proving that the allocators step using related actions ensures that the source and target allocators are in related, consistent states.
This is key to the memory safety preservation part of the theorem, i.e., the right of \Cref{fig:proof}.

To prove memory safety preservation for \Cref{thm:comp-rs-strong}, we start from the functional correctness square between allocators (i.e., the base of the prism on the right in \Cref{fig:proof}).
Then, we assume that the \clang step is memory-safe, this is represented by the blue side of the prism.
Technically, this relies on another omitted piece of formalisation that relates source allocator states \src{A} and shadow memories \com{T}, a relation that holds when the addresses tracked in \src{A} and \com{T} are the same up to a bijection $\srcdelta{}$.
The source memory safety assumption tells that the states of the initial \clang allocator and of the initial shadow memory are related ($\src{A} =_{\srcdelta{}} \com{T}$), that they take a related step ($\as=_{\srcdelta{'}}\ac$), and that leads to related final states ($\src{A'} =_{\srcdelta{'}} \com{T'}$).
The goal of the memory safety part of the proof is depicted as the corresponding red side of the prism: there is a relation between the states of the initial \mswasm allocator and of the initial shadow memory ($\trg{A} =_{\trgdelta{}} \com{T}$), the states take a related step ($\at=_{\trgdelta{'}}\ac$) and that leads to related final states ($\trg{A'}=_{\trgdelta{'}} \com{T'}$).
%
%
%
%
To construct the relations in the red square, we need to derive the dashed edges of the vertical triangles according to the correct relation with the correct bijection.
This relation we obtain by combining the corresponding source-to-target relations (i.e., $\src{A} =_{\delta} \trg{A}$) and the source-to-monitor relation (i.e, $\src{A} =_{\srcdelta{}} \com{T}$), and compose their bijections to relate abstract and target locations.
That is, we obtain $\trg{A} =_{\trgdelta{}} \com{T}$, where $\trgdelta{}$ (relating \mswasm and abstract addresses) is the composition of $\delta$ (relating \mswasm and \clang addresses) with $\srcdelta{}$ (relating \clang and abstract addresses).
Importantly, the triangle of relations guarantees that the \clang notion of memory safety is preserved \emph{exactly} in \mswasm.
Since in \clang we instantiate our abstract notion of memory safety to account for intra-object safety, we get the same fine-grained memory safety notion preserved in \mswasm.



The proof of \Cref{thm:trap} is analogous.
There, we use the same intuition presented above to simulate all actions of the memory-safe trace \alst starting from their memory-safe counterparts in \alss.
Then, at some point, starting from related states, \clang performs a memory-unsafe action \as and \mswasm emits a \trapt.
%
This proof is by case analysis over \clang memory safety violations, which we identify by the related abstract monitor getting stuck.
In the proof, we relate these violations to a \emph{failing} memory safety check in \mswasm, which causes the compiled program to $\trapt$, as expected.
%
%



\section{Implementing \mswasm}\label{sec:impl}

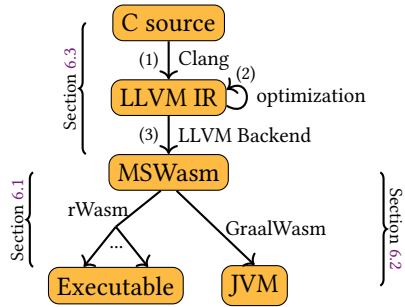
\begin{wrapfigure}[19]{R}{.55\textwidth}
\begin{minipage}{0.53\textwidth}

  \centering
\begin{tikzpicture}
	\tikzstyle{state} = [fill=Dandelion,rounded corners,draw]
	\tikzstyle{nodel} = [midway,left,font=\footnotesize]
	\tikzstyle{noder} = [midway,right,font=\footnotesize]
	\tikzstyle{nodestage} = [font=\scriptsize]

	\begin{pgfonlayer}{foreground}
		\node[state](c){C source};
		\node[state, below =.5 of c](llvm){LLVM IR};
		\node[state, below =.5 of llvm](mswasm){\mswasm};
		\node[state, below =1 of mswasm, xshift=-2em](ex){Executable};
		\node[state, right =.5 of ex](jvm){JVM};

		\draw[->,thick] (c.south) to node[noder](cla){Clang} node[nodestage,left](o){(1)} (llvm.north);
		\draw[->,thick,out=-30,in=30,looseness=5] ([yshift=-.3em]llvm.east) to node[noder](opt){optimization} node[nodestage,pos=.6,above](t){(2)} ([yshift=.3em]llvm.east);
		\draw[->,thick] (llvm.south) to node[noder,align=center](llvmb){LLVM Backend} node[nodestage,left](tt){(3)} (mswasm.north);
		\draw[-,thick] ([xshift=-.3em]mswasm.south) to node[nodel](rw){\rwasm} ([yshift=1.5em]ex.north);
		\draw[->,thick] ([yshift=1.5em]ex.north) to node[noder](t1){...} (ex.150);
		\draw[->,thick] ([yshift=1.5em]ex.north) to node[noder](t2){} (ex.30);

		\draw[->,thick] ([xshift=.3em]mswasm.south) to 
		node[noder](t4){\graalwasm} (jvm.north);


    \draw[decorate,decoration={brace},thick] ([xshift=-1em]c.west|-mswasm.north) to node[nodestage,above,sloped](a){\Cref{sec:comp-impl}} ([xshift=-1em]c.west);
    \draw[decorate,decoration={brace},thick] ([xshift=-3em]c.west|-ex.north) to node[nodestage,above,sloped](b){\Cref{sec:mswasm-impl}} ([xshift=-3em]c.west|-mswasm.center);
    \draw[decorate,decoration={brace},thick] ([xshift=3.5em]jvm.east|-mswasm.center) to node[nodestage,above,sloped](c){\Cref{sec:mswasm-impl-graal}} ([xshift=3.5em]jvm.east|-jvm.center);

	\end{pgfonlayer}
\end{tikzpicture}

  \caption{End-to-end compilation pipeline. We first compile C to \mswasm (via
  LLVM), and then compile \mswasm to machine code using either our modified
  \rwasm AOT compiler (which supports different notions of safety) or our
  modified \graalwasm JIT compiler.}
\label{fig:sys-arch}
\end{minipage}
\end{wrapfigure}

In this section we describe our prototype \mswasm compilation framework
(\Cref{fig:sys-arch}).
We implement two compilers \emph{of} \mswasm following the language semantics
of \Cref{sec:mswasm-lang}.
Our first compiler is an ahead-of-time (AOT) compiler from \wasm to executable
machine code (\Cref{sec:mswasm-impl}); it demonstrates \mswasm's
flexibility in employing different enforcement techniques.
Our second compiler is a compiler from \wasm to Java bytecode
(\Cref{sec:mswasm-impl-graal}); it demonstrates \mswasm's compatiblility with
just-in-time (JIT) compilation.
We also implement a compiler from C \emph{to} \mswasm (\Cref{sec:comp-impl}),
following the formal compiler model of \Cref{sec:compiler}.
We describe these prototypes next.

Our prototype implementation of \mswasm extends the bytecode of \wasm with instructions to manipulate the segment memory as well as handles.
In doing so, it takes a few shortcuts in the name of
expediency---most notably, it replaces the existing \wasm opcodes for \loadt\ and
\storet with \hload and \hstore.
A production \mswasm implementation would
support both segment-based and linear-memory-based operations simultaneously,
by using two-byte opcode sequences for \hload and \hstore.

%

%

\subsection{Ahead of Time Compilation of \mswasm}\label{subsec:rwasm}\label{sec:mswasm-impl}

To compile \mswasm bytecode to machine code, we build on the \rwasm
compiler~\cite{rwasm}.
\rwasm is a provably-safe sandboxing compiler from \wasm to Rust, and thus to
high-performance machine code.\footnote{
  In modifying \rwasm, we were careful to ensure that we preserve its
  previously-established sandboxing/isolation guarantees.
  These guarantees, together with the internal memory-safety guarantees from
  \mswasm, increases the level of protection for native code generated by \rwasm.
}
We extended \rwasm to support \mswasm as follows.
We modified \rwasm's frontend to parse \mswasm instructions
and propagate them through to later phases.
We updated \rwasm's stack analysis to account for \mswasm's new types and
instructions (\eg \hload and \hstore, which take a \handle as argument).
Finally, we updated \rwasm's backend---the code generator, specifically---to
implement \mswasm's instructions and segment memory.

One of the benefits of \mswasm is that it gives \wasm compilers and runtimes
flexibility in how to best enforce memory safety.
This is especially important today: memory-safety hardware support is only
starting to see deployment and applications have different security-performance
requirements---we cannot realistically expect everyone to pay the cost of
software-based memory safety.
When hardware becomes available, \mswasm programs can take advantage of
hardware acceleration almost trivially: in our AOT compiler, for example, we
only need to tweak the codegen stage.
We demonstrate this flexibility by prototyping two different software
techniques that have different safety and performance characteristics.
Future \mswasm implementations could instead use novel hardware extensions such
as Arm MTE~\cite{armmte} or CHERI~\cite{arm-cheri, cheri}.


\myparagraph{Segments as Vectors.}
Our default technique for memory-safety enforcement closely matches
\Cref{sec:sem-intra-pre}, and enforces spatial safety, temporal safety, and
handle integrity (\rwasmdefault in \Cref{sec:eval}).
We implement the segment memory as a vector (\texttt{Vec}) of segments. Each
segment is a pair composed of a \texttt{Vec} of bytes (giving us spatial safety)
and a \texttt{Vec} of tags, which is used to enforce
handle integrity. Handles themselves are implemented using a tagged
\texttt{enum}. To enforce temporal safety we clear free segments from memory
and use sentinel value to prevent the reuse of segment indexes.
A slight variation of this technique (\rwasmnohi in \Cref{sec:eval}) gives up
on handle integrity (we remove the \texttt{Vec} of tags and related checks)
for performance.
%

\myparagraph{Segments with Baggy Bounds.}
Our second technique is inspired by Baggy Bounds checking~\cite{baggy-bounds},
which performs fast checks at each handle-modifying operation and elides checks
at loads and stores. This technique gives up on handle integrity and temporal
safety, since accesses are not checked, but is considerably faster (\rwasmbaggy
in \Cref{sec:eval}).
To implement this technique, our compiler uses a single growable \texttt{Vec}
of bytes, within which a binary buddy allocator allocates implicit segment
boundaries.
We implement the handles as 64-bit values storing an offset in memory and the
$\log$ of the segment size (rounded up to nearest power of two at allocation).
We emit bounds checks for each operation that might modify handles, ensuring
that handles remain within the (baggy) bounds of their corresponding segment.
Specifically, when handles stray a short distance outside their segment, we
mark them as such (and they can safely return back), but we trap when
they (try to) stray too far.

\myparagraph{Implementation Effort.}
Our modifications to \rwasm comprise roughly \locrwasm lines of
additional code,
including both memory-safety enforcement techniques.
The implementation of these two techniques comprise approximately 500
lines of code each in \rwasm's codegen, and share the
rest of \rwasm's codebase.
The relative ease of these modifications illustrates how \mswasm provides a
fertile ground for experimenting with new techniques for providing
performant memory safety.

\subsection{Just in Time Compilation of \mswasm}\label{subsec:graalwasm}\label{sec:mswasm-impl-graal}

Our second prototype is a just-in-time compiler of \mswasm built on top of
\graalwasm~\cite{graalwasm}.
\graalwasm is a \wasm frontend for \graalvm~\cite{graalvm},
a JVM-based JIT compiler capable of compiling a wide range of languages through
the Truffle framework \cite{graalvm-truffle}.
We extend \graalwasm to support \mswasm.
Our modifications mirror those we made to \rwasm:
We modified the \graalwasm frontend to parse \mswasm and the backend---the
\graalwasm interpreter in this case---to support \mswasm's instructions and
segment memory model.
We were able to reuse the \graalvm JIT compiler unmodified, as it automatically
optimizes the AST generated by Truffle from the interpreter.

\myparagraph{Segments as Objects.}
Unlike our \rwasm implementation, we only consider one enforcement technique.
We pick a middle ground between safety and performance: We enforce spatial and
temporal safety, but not handle integrity (\graalwasmmswasm in
\Cref{sec:eval}).
Our implementation of memory segments in \graalwasm is similar to our first
\rwasm technique (but does not track handle-integrity tags).
We implement the segment memory as a Java object, \texttt{SegmentMemory}, which
tracks a list of segments.
\texttt{SegmentMemory} is backed by Java's \texttt{Unsafe} memory manager, and
creates new segments by manually allocating a new chunk of \texttt{Unsafe}
memory.
A segment is represented by a \texttt{Segment} object, which contains an
address within the \texttt{Unsafe} memory, the (inclusive) upper bound of the
segment in memory, and a randomly generated key.
%
%
To ensure temporal safety, free segments are removed from the list of segments
in \texttt{SegmentMemory}, leaving no way to reference them.

\myparagraph{Implementation Effort.}
We added roughly \locgraal lines of code to \graalwasm.
Our prototype is relatively simple and not yet tuned to take full advantage of
\graalvm's optimizations.
We leave this to future work.

\subsection{Compiling C to \mswasm}\label{subsec:c-to-mswasm}\label{sec:comp-impl}
\mswasm, like \wasm, is intended to be a compilation target from higher level
languages.
We implement a compiler from C to \mswasm by extending the CHERI fork of
Clang and LLVM~\cite{cheri-llvm}.
CHERI modified LLVM to support fat pointers, which share many characteristics
with \mswasm handles, and is thus a good starting point for \mswasm.

CHERI represents fat pointers at the LLVM IR level as pointers in a
special, distinguished ``address space''; pointers in this address space are
lowered to CHERI capabilities in the appropriate LLVM backends.
CHERI today only targets MIPS and RISC-V (with CHERI hardware extensions)
backends; other backends, including the \wasm backend, are incompatible with
CHERI's fat pointers.
We modified the \wasm backend to emit \mswasm bytecode, lowering fat-pointer
abstractions to \mswasm abstractions.
Since most of the implementation details follow from \Cref{sec:compiler}, we
focus on details not captured by our formal model.

\myparagraph{Global and Static Data.}
Our C-to-\mswasm compiler only emits \handle-based load and store operations,
resulting in \mswasm programs which do not use the linear memory at all.
This provides additional safety guarantees (and implementation expediency)
at the expense of some flexibility (\eg we do not support integer-to-pointer
casts, except for a few special cases like constant \texttt{0}).
One consequence of this is that even global variables and static data need to
be accessed via \handles, and thus placed in the segment memory.\footnote{
More precisely, global variables which the program never takes the address of,
do not need this treatment, as we can compile them into \wasm globals; but
global variables which the program does take the address of, such as global
arrays, are accessed via pointers and thus must be located in the segment
memory.}
Our compiler emits instructions to allocate a segment for each LLVM global
variable and store the corresponding \handle in a \wasm global variable.
When the target program needs a pointer to the global array, it simply retrieves
the \handle from the appropriate \wasm global variable.

Some global variables in C are themselves pointers, initialized via
initialization expressions, and need to be pointing to valid, initialized
memory at the beginning of the program.
Our compiler generates the necessary information in the output
\texttt{.wasm} file to instruct \mswasm compilers and runtimes (e.g., \rwasm
and \graalwasm) to initialize certain segments at module initialization time.
%

\myparagraph{C Stack.}
We compile part of the C stack to the segment memory.
Specifically, stack variables whose address-of are taken and stack-allocated
arrays cannot be placed on the (simple and safe) \wasm stack.
Compilers from C to ordinary \wasm place these variables in the linear memory;
our compiler places them in the segment memory.\footnote{
Stack variables which the program never takes the address of can be compiled
to \wasm local variables, and data such as return addresses are never placed
in the linear memory at all; \wasm implementations place them on a safe stack
which is inaccessible to \wasm load and store instructions. The only stack
variables which need to be placed in the linear memory, or for us the segment
memory, are those we need pointers to.}
We allocate a single large segment to represent stack memory for all of the
variables which must be allocated in the segment memory; this means we have a
single stack pointer, which we store in a dedicated \wasm global variable of
type \handle.
Compared to using a separate segment for each stack allocation, our
single-segment implementation is simpler (and faster) but trades-off some
safety, e.g., we cannot prevent a stack buffer overflow from
corrupting another stack-allocated buffer.

\myparagraph{Standard library.}
\wasm programs which depend on \libc need a \wasm-compatible implementation of
\libc.
We modified WASI~\cite{wasi} to be compatible with \mswasm to the extent
necessary for our benchmarks.
Most importantly, we fully recompiled the WASI \libc using our \mswasm compiler,
in order to generate \libc bytecode compatible with \mswasm.
In our \mswasm version of the WASI \libc, the implementations of \texttt{malloc}
and \texttt{free} are completely replaced by trivial implementations consisting
of the \newst\ and \freest\ \mswasm instructions.


\myparagraph{Implementation Effort.}
Our CHERI LLVM additions (in particular to its \wasm backend) and the
WASI \libc, amounted to approximately \locllvm lines of code.
While our compiler can target any \mswasm backend, compiling general, real-world
applications would likely require additional changes to WASI \libc.
We leave this to future work.

\section{Performance Evaluation}\label{sec:eval}
In this section we describe our performance evaluation of the \mswasm compiler.
We use the PolyBenchC benchmarking suite~\cite{polybench-c} since PolyBenchC
has become the de-facto suite used by almost all \wasm compilers.
We compare the performance of \mswasm to the performance of the same benchmarks
compiled to normal \wasm, on each of our implementations.

\myparagraph{Machine setup.}
We compile all benchmarks from C to \wasm using Clang, and from C to \mswasm
using our modified CHERI Clang compiler; in both cases we set the optimization
level to \texttt{-O3}.
We run all benchmarks on a single core on a Linux-based system with an Intel
Xeon 8160.



\begin{figure}[ht]
  \centering
  \includegraphics[width=.5\columnwidth]{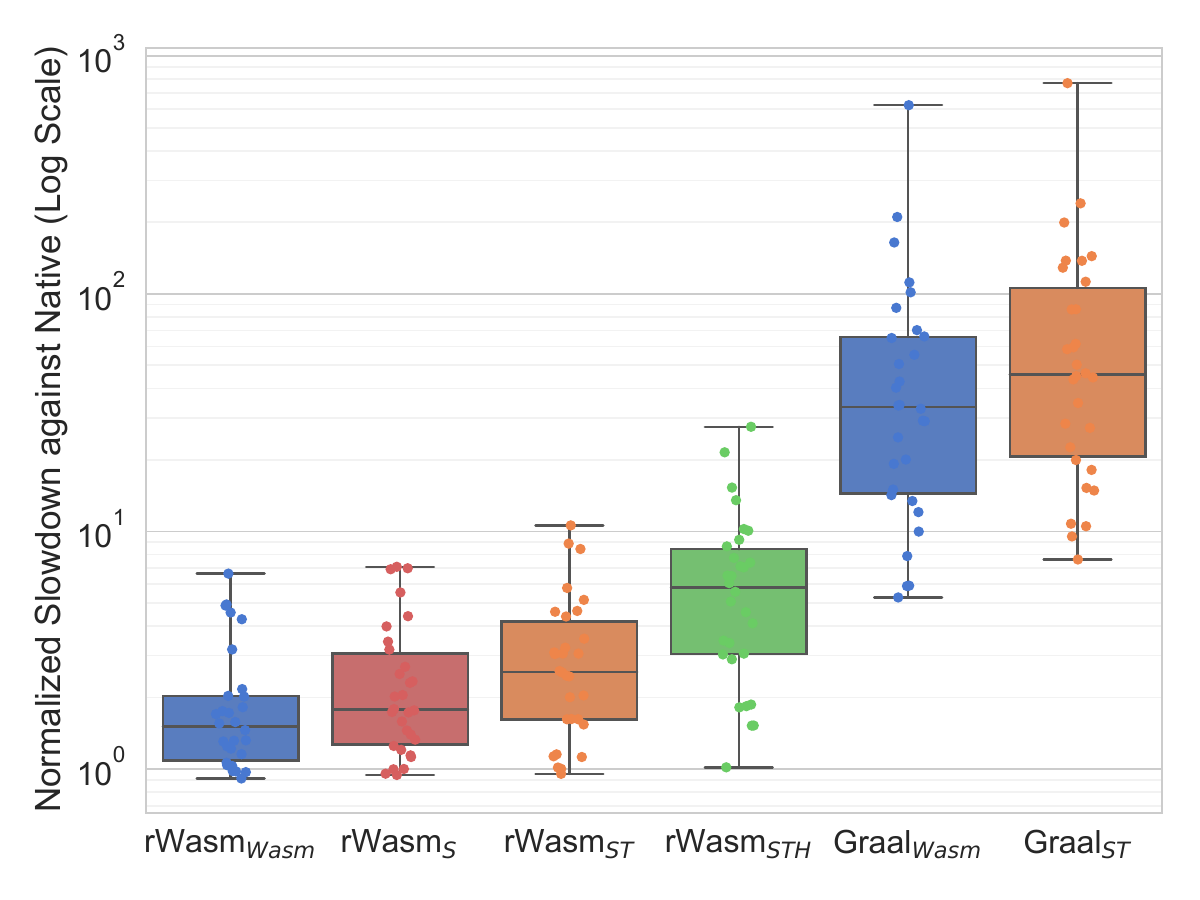}
  \caption{Performance of our implementations of \mswasm compared to
    normal \wasm, normalized against native (non-\wasm) execution on
    benchmarks from PolyBenchC}
  \label{fig:exec-time}
\end{figure}


\myparagraph{Results.}
\Cref{fig:exec-time} summarizes our measurements, normalized against the
execution time of native (non-Wasm) execution.
In this figure, \rwasmnormalwasm and \graalwasmnormalwasm refer to
execution of normal \wasm.
We distinguish the different \mswasm compilers according to their enforcement
techniques: \rwasmdefault enforces spatial safety, temporal safety, and handle
integrity; \rwasmnohi and \graalwasmmswasm only enforce enforce spatial and
temporal safety; and, \rwasmbaggy only enforces spatial safety (in the style of
baggy bounds).
 
As expected, and in line with prior work~\cite{cets,softbound}, each
safety enforcement techniques comes with a performance cost---handle integrity
being the most expensive.
For the AOT compiler, we observe that enforcing spatial safety alone
\rwasmbaggy has a geomean overhead of \evalrwasmbaggyovernormal over
\rwasmnormalwasm; additionally enforcing temporal safety (\rwasmnohi) results
in an overhead of \evalrwasmnohiovernormal over \rwasmnormalwasm; and,
finally, further enforcing handle integrity (\rwasmdefault) increases the
end-to-end overhead to \evalrwasmdefaultovernormal.
For the JIT compiler, enforcing spatial and temporal safety results in an
overhead comparable to that of the AOT compiler: \graalwasmmswasm imposes a
\evalgraalmswasmovernormal geomean overhead.
The JIT approach is much slower than the AOT approach though---the overheads
of \rwasmnormalwasm and \graalwasmnormalwasm over native (non-\wasm) execution
are \evalrwasmnormalovernative and \evalgraalnormalovernative respectively.
We also note that with increasing iterations of the \graalvm JIT,
\graalwasmnormalwasm's performance improves more rapidly than
\graalwasmmswasm's, which suggests that our implementation still has potential
to make better use of \graalvm's optimizer.

Since normal \wasm and \mswasm have different bytecode formats, our evaluation
of \mswasm performance necessarily includes slowdowns caused by inefficiencies
in our compilation from C to \mswasm.
But because \mswasm decouples memory safety enforcement from the generation of
\mswasm bytecode, both parts of this pipeline (C-to-\mswasm compilation, and
\mswasm to machine code) can be independently optimized, with \mswasm performance
benefiting from improvements on both sides.


\section{Related Work}\label{sec:rw}
\myparagraph{Memory safety for C-like languages.}
%
%
Despite a tremendous amount of work on memory-safety protection
mechanisms~\cite{Szekeres:2013}, researchers still struggle to agree
on a common definition for \emph{memory
safety}~\cite{what-is-ms-hicks}.
Azevedo de \citep{Amorim18} characterize memory safety as a 2-hypersafety
property, similar to non-interference.
Their definition belongs to a richer class of security properties,
which are harder to enforce and to preserve robustly through compilation~\cite{rhc}.
%
%
%
%
%

%
Many compiler-based instrumentations have been proposed to enforce
memory safety in C programs via software-based checks attached to
pointer and memory
operations~\cite{softbound,checkedc,ccured,safec,270632,mscc,10.5555/250900.250910,baggy-bounds}.
Some of these solutions are also supported by formal memory-safety
guarantees~\cite{cets,checkedc,softbound}.
These formal results however, are not \emph{robust}, i.e., they do not
guarantee memory safety when linking with arbitrary adversarial code.
Moreover, these formalizations do not actually include the
instrumentation pass of the compiler, but prove memory safety via
\emph{type safety} of an instrumented C-like language, where pointers
are annotated with bounds metadata.
Unlike \mswasm, these languages adopt a high-level memory model, which
implicitly provides pointer integrity.
%
%

%
Our color-based memory-safety monitor and similarly our notion of
authentic pointers and handles are inspired by previous work on
\emph{pointer provenance} in C~\cite{pointer-provenance}.
Some of the C semantics proposed in that work track pointer provenance
also through integer and pointer casts, which we do not consider in
this work, also given that \mswasm has no native notion of casts.
%
%




\myparagraph{Efficient memory-safety implementations.}
%
%
Unlike compiler-based instrumentations, compiling to \mswasm does not
commit to a particular concrete strategy for enforcing memory safety:
Different implementations of \mswasm can use different enforcement
approaches.
In particular, \mswasm enables backends compilers and runtimes to
leverage efficient software- and hardware-based mechanisms,
independently proposed to enforce pointer integrity~\cite{pac},
spatial~\cite{armmte,deltapointers,baggy-bounds}, and
temporal~\cite{parkinson2017project,DangNull} safety, to create new
practical memory-safety enforcement schemes.
Because \mswasm is platform-agnostic, we expect that implementations
will be able to opportunistically take advantage of hardware memory
protection mechanisms on individual
platforms~\cite{armmte,10.1145/3224423,10.1145/1346281.1346295,10.1145/2508859.2516713}
(current and proposed) to efficiently implement handles.
If CHERI~\cite{cheri} becomes widespread, we expect that extant
\mswasm code would be able to directly take advantage of its hardware
capabilities, something that would not be possible for native binaries
previously compiled using a software-based protection scheme.
%

\myparagraph{Software isolation via \wasm.}
%
%
\wasm abstractions provide an efficient software-isolation mechanism,
which has been applied in many different domains.
For example, using \wasm, the RLBox framework~\cite{rlbox} retrofits
isolation into the Firefox browser; Sledge~\cite{sledge} enables
lightweight serverless-first computing on the Edge;
and eWASM~\cite{ewasm} demonstrates practical software fault
isolation for resource-constrained embedded platforms.
These use cases already rely on both the performance and the sandboxing safety
of \wasm, and stand to benefit from \mswasm's focus on memory safety.

\citep{rwasm} use formal methods and non-traditional techniques
respectively to provide provable isolation between the Wasm module,
running as a native library, and the host process executing
it.
Their focus is on provable module--host isolation, and module-internal
memory safety is explicitly left out of scope.
As shown by \citep{Lehmann20}, \wasm lacks many common
defenses (e.g., stack canaries, guard pages, ASLR) against classic
memory safety vulnerabilities, such as buffer overflows.
%
%


\citep{notsofast} perform a large-scale performance evaluation of
browser \wasm runtimes, comparing to native code.
%
%
Our evaluation of \mswasm's performance (\Cref{sec:eval}) shows that adding
memory-safety protections does not fundamentally change \wasm's performance
story.
In particular, adding spatial and temporal safety imposes less overhead on
\wasm than the overhead \wasm already incurs vs native code.

\section{Conclusion}\label{sec:conc}
This paper realised the \mswasm proposal to extend \wasm with language-level memory-safety abstractions, giving it a formal semantics, proving that its programs are all memory safe and implementing the \mswasm language runtime.
Like \wasm, \mswasm is intended to be used as a compilation target, so this paper formalised a \clang-to-\mswasm compiler, proved that it enforces memory safety, and implemented variations of said compiler with different tradeoffs between speed and security.
%
%
Our PolyBenchC-based evaluation shows that \mswasm introduces an overhead ranging from 22\% (enforcing spatial safety alone) to 198\% (enforcing full memory safety).
Our software-based implementations only serve to highlight that enforcing
memory safety for \wasm is possible and, moreover, that \mswasm makes it easy
to change the underlying enforcement mechanism without modifying application
code.
This means \mswasm engines will be able to take advantage of clever memory
safety enforcement techniques today and hardware extensions in the near future,
progressively (and transparently) improving the safety of the applications they
run.

\begin{acks}
  This work was partially supported:
    by the German Federal Ministry of Education and Research (BMBF) through funding for the CISPA-Stanford Center for Cybersecurity (FKZ: 13N1S0762),
    by the Italian Ministry of Education through funding for the Rita Levi Montalcini grant (call of 2019). Gollamudi was supported in part through a generous gift to support research on applied cryptography and society in the Center for Research on Computation and Society (Harvard University) and Computing Innovating Fellowship (2021).
\end{acks}



%
%
%
%


\newpage
\bibliography{biblio.bib}

\end{document}